%% file: ms.tex
\documentclass[times, twocolumn]{aastex63}
\usepackage{CJK}
\usepackage{graphicx}
\usepackage{enumerate}
\usepackage{amssymb, amsmath}
\usepackage{natbib}
\usepackage{color}
\usepackage{ulem}
\usepackage{appendix}
\usepackage{hyperref}

\usepackage[nodayofweek]{datetime}
\newdateformat{monthyearday}{%
  \THEYEAR\ \monthname[\THEMONTH] \twodigit{\THEDAY}}

\newcommand{\teff}{\ensuremath{T_{\rm eff}}}
\newcommand{\rsun}{\ensuremath{\,R_\Sun}}
\newcommand{\lsun}{\ensuremath{\,L_\Sun}}
\newcommand{\msun}{\ensuremath{\,M_\Sun}}
\newcommand{\mj}{\ensuremath{\,M_{\rm J}}}

\newcommand{\mum}{$\mu$m}
\newcommand{\hd}{HD\,19467}
\newcommand{\hdb}{HD\,19467\,B}
\newcommand{\um}{$\rm\,\mu m$~}

\input{commands-star}

\begin{document}
\accepted{January 25, 2023}

\shorttitle{JWST Observations of \hdb}
\shortauthors{Greenbaum et al.}

\title{First Observations of the Brown Dwarf \hdb\  with JWST}

\correspondingauthor{Alexandra Z. Greenbaum}
\email{azg@ipac.caltech.edu}

\author[0000-0002-7162-8036]{Alexandra Z. Greenbaum}
\affiliation{IPAC, Caltech, 1200 E. California Blvd., Pasadena, CA 91125, USA}

\author[0000-0002-3414-784X]{Jorge Llop-Sayson}
\affiliation{California Institute of Technology, 1200 E. California Blvd., Pasadena, CA 91125, USA}
\author[0000-0003-1487-6452]{Ben Lew}
\affiliation{NASA Ames Research Center, Mountain View, CA, 94035, USA}
\author[0000-0001-5966-837X]{Geoffrey Bryden}
\affiliation{Jet Propulsion Laboratory, California Institute of Technology, Pasadena, CA 91109, USA}
\author {Thomas Roellig}
\affiliation{NASA Ames Research Center, Mountain View, CA, 94035, USA}
\author[0000-0001-7591-2731]{Marie Ygouf}
\affiliation{Jet Propulsion Laboratory, California Institute of Technology, Pasadena, CA 91109, USA}
\author[0000-0003-3504-5316]{B.J. Fulton}
\affiliation{NASA Exoplanet Science Institute, Caltech, 1200 E. California Blvd., Pasadena, CA 91125, USA}
\author[0000-0001-8832-4488]{Daniel R.\ Hey}
\affiliation{Institute for Astronomy, University of Hawai`i, 2680 Woodlawn Drive, Honolulu, HI 96822, USA}
\author[0000-0001-8832-4488]{Daniel Huber}
\affiliation{Institute for Astronomy, University of Hawai`i, 2680 Woodlawn Drive, Honolulu, HI 96822, USA}
\author[0000-0003-1622-1302]{Sagnick Mukherjee}
\affiliation{Department of Astronomy and Astrophysics, University of California, Santa Cruz, CA 95064, USA}
\author[0000-0003-1227-3084]{Michael Meyer}
\affiliation{Department of Astronomy, University of Michigan, Ann Arbor, MI 48109, USA}
\author[0000-0002-0834-6140]{Jarron Leisenring}
\affiliation{Steward Observatory, University of Arizona, Tucson, AZ 85721, USA}
\author[0000-0002-7893-6170]{Marcia Rieke}
\affiliation{Steward Observatory, University of Arizona, Tucson, AZ 85721, USA}
\author{Martha Boyer}
\affiliation{Space Telescope Science Institute, 3700 San Martin Drive, Baltimore, MD 21218, USA}
\author{Joseph J. Green}
\affiliation{Jet Propulsion Laboratory, California Institute of Technology, Pasadena, CA 91109, USA}
\author{Doug Kelly}
\affiliation{Steward Observatory, University of Arizona, Tucson, AZ 85721, USA}
\author{Karl Misselt}
\affiliation{Steward Observatory, University of Arizona, Tucson, AZ 85721, USA}
\author{Eugene Serabyn}
\affiliation{Jet Propulsion Laboratory, California Institute of Technology, Pasadena, CA 91109, USA}
\author{John Stansberry}
\affiliation{Space Telescope Science Institute, 3700 San Martin Drive, Baltimore, MD 21218, USA}
\author[0000-0002-1437-4463]{Laurie E. U. Chu}
\affiliation{NASA Postdoctoral Program Fellow, NASA Ames Research Center, M/S 245-1, Moffett Field, CA 94035, USA}
\author[0000-0003-1863-4960]{Matthew De Furio}
\affiliation{Department of Astronomy, University of Michigan, Ann Arbor, MI 48109, USA}
\author[0000-0002-6773-459X]{Doug Johnstone}
\affiliation{NRC Herzberg Astronomy and Astrophysics, 5071 West Saanich Rd, Victoria, BC, V9E 2E7, Canada}
\affiliation{Department of Physics and Astronomy, University of Victoria, Victoria, BC, V8P 5C2, Canada}
\author{Joshua E. Schlieder}
\affiliation{Exoplanets and Stellar Astrophysics Laboratory, NASA Goddard Space Flight Center, 8800 Greenbelt Road, Greenbelt, MD, USA}

\author[0000-0002-5627-5471]{Charles Beichman}
\affiliation{Jet Propulsion Laboratory, California Institute of Technology, Pasadena, CA 91109, USA}
\affiliation{NASA Exoplanet Science Institute, Caltech, 1200 E. California Blvd., Pasadena, CA 91125, USA}

\begin{abstract}
We observed \hdb\ with JWST's NIRCam in six filters spanning 2.5-4.6 \mum\ with the Long Wavelength Bar coronagraph. The brown dwarf \hdb\ was initially identified through a long-period trend in the radial velocity of G3V star \hd. \hdb\ was subsequently detected  via coronagraphic imaging  and spectroscopy, and characterized as a late-T type brown dwarf with approximate temperature $\sim1000$~K. We observed \hdb\ as a part of the NIRCam GTO science program, demonstrating the first use of the NIRCam  Long Wavelength Bar coronagraphic mask. The object was detected in all 6 filters (contrast levels of $2\times10^{-4}$ to $2\times10^{-5}$) at a separation of 1.6\arcsec\ using  Angular Differential  Imaging (ADI) and Synthetic Reference Differential Imaging (SynRDI).
Due to a guidestar failure during acquisition of a  pre-selected reference star, no reference star data was available for post-processing. However, RDI was successfully applied using synthetic Point Spread Functions (PSFs) developed from contemporaneous  maps of the telescope's optical configuration.   Additional radial velocity data  (from Keck/HIRES) are  used to  constrain the orbit of the \hdb. Photometric data from TESS are used to constrain the properties of the host star, particularly its age. NIRCam photometry, spectra and photometry from literature, and improved stellar parameters  are used in conjunction with recent spectral and evolutionary substellar models to derive physical properties for \hdb. Using an age of 9.4$\pm$0.9 Gyr inferred from spectroscopy, Gaia astrometry, and TESS asteroseismology, we obtain a model-derived mass of 62$\pm 1$\mj, which is consistent within 2-$\sigma$ with the dynamically derived mass of 81$^{+14}_{-12}$\mj.
\end{abstract}

\section{Introduction}
Brown dwarfs provide a unique testbed for confronting evolutionary and atmospheric models of sub-stellar objects with well-defined observations. Those brown dwarfs which are companions to main sequence stars, as opposed to free-floating, are particularly valuable since they are presumed to inherit observable stellar properties such as metallicity and share similar ages. This knowledge constrains many of the free parameters in the comparison of models  with observation.  

Low-mass brown dwarf companions to main-sequence stars were initially found through blind imaging searches, e.g. GL229 B \citep{Nakajima1995}, and subsequently as a by-product of planet searches using the radial velocity (RV) technique. In the case of \hd, \citet{Crepp2014} identified it as a star with a significant RV trend suggestive of a massive brown dwarf companion. 
Coronagraphic imaging with Keck NIRC2 first confirmed the presence of the companion \citep{Crepp2014}. This was followed by spectroscopy with Palomar's P1640 instrument that characterized \hdb \ as a brown dwarf with effective temperature of $\sim$1000 K corresponding to a T5.5 spectral type \citep{Crepp2015}. More recently, proper motion measurements from the Hipparcos and Gaia catalogs have been used to identify systems with companions or help characterize them, including \hd\ \citep{Brandt2021}. 

Multiple JWST programs will provide imaging and spectroscopy of \hdb\ across the near- and mid-IR where brown dwarfs emit most of their energy. The program presented here (PID \#1189) uses NIRCam \citep{Rieke2022} to provide medium and narrow band imaging and photometry of \hdb\ in 6 bands, spanning 2.5 to 4.5 \mum. At a later date, another JWST program (PID \#1414) will use NIRSpec \citep{Jakobsen2022} to obtain high-resolution (R $\simeq$ 2700) 3--5 \mum\ spectra of \hdb. 
 
JWST observations of the G3V star \hd\ with its T5+ brown dwarf companion, \hdb\ \citep{Crepp2014}, represent one of the earliest exercises of the NIRCam Coronagraphic LW Bar \citep{Krist2007, Beichman2010, Girard2022}, providing an opportunity for an early scientific result  and a demonstration of the capabilities of the instrument.

The NIRCam observations presented in this study are designed to accomplish three main goals: 
\begin{enumerate}
    \item Provide an early dataset that exercises the bar mask on NIRCam, especially without a reference star (\S\ref{sec:observations} \& \ref{sec:NIRCAMresults});
    \item Refine the orbital parameters of \hdb\ with a new imaging data point along with  new RV data from Keck/HIRES (\S \ref{sec:orbitFitting}); and
    \item Add additional photometric measurements to better constrain the physical properties of the brown dwarf (\S \ref{sec:atmosphereFitting}).
\end{enumerate}
We also include new analysis of TESS observations to constrain properties of the host star (\S \ref{sec:hostStar}).

\section{Observations}\label{sec:observations}
\subsection{NIRCam Observations}
NIRCam observed \hd\ on 2022-Aug-12 with the long-wavelength bar (LWB) coronagraphic mask in subarray mode with six filters: F250M, F300M, F360M, F410M, F430M, and F460M.  The target star was observed at two telescope roll angles separated by 7.72 degrees.
Table \ref{tab:exposures} shows a summary of the observations  and settings per filter.
Observations of \hd\ were taken with the long-wavelength bar (MASKLWB) coronagraph, providing a test of NIRCam's capabilities  at smaller inner working angles than are possible with the round masks ($4\lambda/D$ for MASKLWB vs $6\lambda/D$ for MASK210R, MASK335R, and MASK430R; \citet{Krist2007}). At the time of these observations, the MASKLWB positions were not well-defined, with a y-offset $\sim$70 mas. Future use of this mode with updated position definition will improve the ability to center the star on the mask and therefore contrast performance, especially close in to the mask. These observations represent one of the first post-commissioning uses of the bar coronagraph. 

The observation plan was initially scheduled to include sequential observations of the reference star HD 19096 in order to perform PSF subtraction using Reference Differential Imaging (RDI). However, 
the reference observations were unsuccessful because the telescope failed to acquire a guide star.
Instead, we performed post-processing using only angular diversity along with models of the telescope and instrument's optical performance enabled by regular measurements of the telescope wavefront error, simulating Reference Star Differential Imaging but without the actual observation of a reference star. A very similar approach has been applied to enable high contrast imaging with the Hubble Space Telescope by modeling the instrument PSF \citep[e.g.][]{Krist1997}; JWST's stability and regular measurements of the wavefront further enable this technique.
High contrast observations with only angular diversity can significantly reduce the observation time and overhead. We demonstrate that this can be an appropriate strategy for bright and widely separated companions.

\begin{deluxetable}{ccccccc}
\tabletypesize{\scriptsize}
\tablewidth{0pt}
\tablecaption{NIRCam Observing Parameters (PID:\#1189)\label{tab:exposures}
}
\tablehead{
\colhead{Target} & \colhead{Filter}& \colhead{Readout} & \colhead{Groups/Int} & \colhead{Ints/Exp}& \colhead{Dithers} & \colhead{Exp Time (s)}}
\startdata
\multicolumn{7}{l}{\textit{Subarray SUB320; Roll 1}}\\
\hd\   &F250M  &MEDIUM2    &10&10&1 &983.517 \\
\hd\  &F300M  &MEDIUM2   &10&5&1 &491.758\\
\hd\  &F360M  &MEDIUM2   &10&5&1 &491.758\\
\hd\  &F410M  &MEDIUM2   &10&5&1 &491.758\\
\hd\  &F430M  &MEDIUM2   &10&5&1 &491.758\\
\hd\  &F460M  &MEDIUM2   &10&5&1 &491.758\\
\hline
\multicolumn{7}{l}{\textit{Subarray SUB320; Roll 2}}\\
\hd\  &F250M  &MEDIUM2    &10&10&1 &983.517 \\
\hd\  &F300M  &MEDIUM2   &10&5&1 &491.758\\
\hd\  &F360M  &MEDIUM2   &10&5&1 &491.758\\
\hd\  &F410M  &MEDIUM2   &10&5&1 &491.758\\
\hd\  &F430M  &MEDIUM2   &10&5&1 &491.758\\
\hd\  &F460M  &MEDIUM2   &10&5&1 &491.758\\
\enddata
\tablecomments{Observations of reference star HD19096 were not executed.}
\tablecomments{Total Time refers to the effective exposure time reported in the data headers, keyword XPOSURE.}
\end{deluxetable}

\subsection{Radial Velocity Observations\label{sec:RVfit}}

New radial velocity measurements of \hd\ were obtained in July through August 2022 using the High Resolution spectrometer (HIRES) on the Keck I Telescope. The new RV measurements are processed using standard data reduction techniques described in \citet{Butler1996} and \citet{Butler2017}. The majority of the RVs come from \citet{Rosenthal2021}, where the reduction techniques are described in more detail. In brief, the HIRES RV values are measured using an iodine cell-based design in order to wavelength calibrate the stellar spectrum. The spectral region from 5000-6200 \AA\ is used for measuring the radial velocities. We combine the new observations with previous measurements for a total of 53 RV measurements spanning 25 years for the data analysis. The data including the new measurements are listed in Table~\ref{tab:NEW_RV} in Appendix \ref{sec:RVapp}.

\subsection{TESS Observations \label{sec:TESS}}

\hd\ was observed by the TESS spacecraft \citep{Ricker2015} in Sectors 4 and 31, resulting in $\approx$\,60 days of high-precision optical photometry. Sector 31 includes data obtained with 20-second cadence, a new observing mode introduced in the TESS extended mission. TESS 20-second data shows improved photometric precision for bright stars such as \hd\ \citep{huber22}, and we therefore focus on 20-second data here. We used the PDC-MAP light curves provided by the Science Processing Operations Center \citep[SPOC,][]{jenkins16}, which have been optimized to remove instrumental variability \citep{smith12,stumpe12}, and remove all data with quality flags not equal to zero which yields the best precision for 20-second data \citep{huber22}.

\section{Host Star Properties}\label{sec:hostStar}

\hd\ is a slightly metal poor G3 main sequence star \citep{Gomes2021} as summarized in Table~\ref{tab:star}. In some cases multiple values are given for key parameters to give an idea of their spread. The biggest discrepancy concerns the  age estimates which range from 5.41$^{+1.8}_{-1.34}$ to 11.88$\pm2.56$  Gyr \citep{Brandt2021, Wood2019, Maire2020, Gomes2021}.   \citet{Maire2020} apply different approaches to age determination and provide a thorough discussion of their merits and drawbacks. Their work generally suggests older ages than the initial \citet{Crepp2014} estimation but they note that the chemical abundance and kinematics likely place \hd\ in the thin disk population, suggesting an age younger than 10~Gyr. 
We discuss our choice of age in more detail below, including new asteroseismology data from TESS, which favor an older age.

\subsection{Analysis of TESS photometry}
\begin{figure}
\begin{center}
\includegraphics[width=\linewidth]{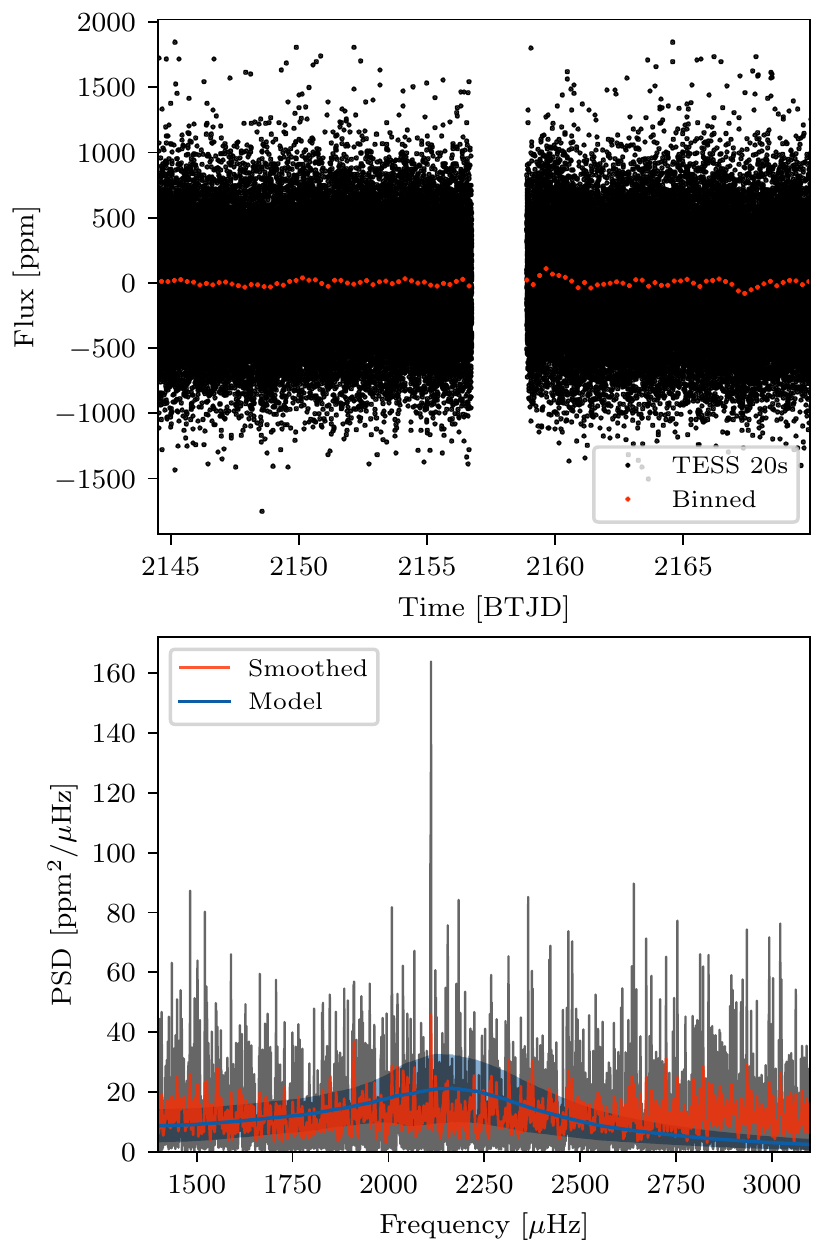}
\caption{Top: TESS Sector 31 light curve of \hd. Black points show the original 20-second cadence data, red points show the data binned to a timescale of 6 hours. Bottom: Power spectrum of the data centered on the detected power excess near $\approx$\,2200 \muHz. The blue line and filled area shows the median and standard deviation of the GP model posterior}
\label{fig:lc}
\end{center}
\end{figure}

The top panel of Figure \ref{fig:lc} shows the TESS 20-second cadence light curve for \hd. We observe no significant long-term variability, with an RMS of 25.5 ppm over 6 hour timescales. To search for high-frequency variability, we used the established asteroseismic tools  \texttt{pySYD} \citep{huber09,chontos21} and \texttt{FAMED} \citep{famed}, which analyze the data in the frequency domain. Both methods detected a significant power excess near $\approx$\,2200 \muHz, consistent with the expected $\approx$7 minute timescale of solar-like oscillations \citep{Bedding2014, Garcia2019} based on the spectroscopic temperature and surface gravity (Table 2).
We also analyzed the data in the time-domain using a Gaussian Process (GP) model with a stochastically driven damped harmonic oscillator \citep{dfm17}, which has been demonstrated to outperform traditional frequency analysis tools in recovering low S/N oscillations (Hey et al., in prep). The GP analysis strongly favored a model with an oscillating component with a $\Delta$BIC=7.1.

The bottom panel of Figure \ref{fig:lc} shows the power spectrum of the 20-second light curve centered on the  power excess. Solar-like oscillations are described by a frequency of maximum power (\numax) and a large frequency separation (\dnu), which approximately scale with \logg\ and the mean stellar density, respectively \citep{ulrich86,brown91}. We derive  $\numax=2180 \pm 100$ $\muHz$, with the central value taken from the median of three solutions (pySYD, FAMED, GP), and uncertainties calculated from the scatter over individual methods \citep[e.g.][]{huber13}. The low S/N of the detection precludes an unambiguous detection of \dnu. Visual inspection of an echelle diagram indicates $\dnu\approx 101$\muHz, consistent with the derived \numax value.

\subsection{Physical Properties of HD 19467} \label{sec:hostStarProperties}

\begin{deluxetable}{llll}[t!]
\tabletypesize{\scriptsize}
\tablewidth{0pt}
\tablecaption{Observations of the Host Star \hd\label{tab:star}}
\tablehead{
\colhead{Property} & \colhead{Value}& \colhead{Units}& \colhead{Comments} }
\startdata
Spectral Type & G3V & &\citet{Gomes2021}\\
T$_{\rm eff}$ &5720$\pm$10 & K &\citet{Gomes2021}\\
T$_{\rm eff}$ &5747$\pm$25 & K &\citet{brewer16}\\
T$_{\rm eff}$ &5770$\pm$80 & K &\citet{Maire2020}\\
T$_{\rm eff}$ &5742$\pm$10 & K &\citet{nissen20}\\\hline
Mass &0.953$\pm$0.022 & M$_\odot$&\citet{Maire2020}\\
Mass &0.960$\pm{0.02}$ & M$_\odot$& This work ($\S$\ref{sec:hostStarProperties})\\\hline
Age &5.41$^{+1.8}_{-1.34}$ &Gyr & \citet{Brandt2021}\\
Age &8.0$^{+2.0}_{-1.0}$&Gyr & \citet[Table 1]{Maire2020} \\ 
Age & 10.06$^{+1.16}_{-0.82}$ & Gyr & \citet{Wood2019} \\
Age &11.882$\pm{2.564}$&Gyr & \citet{Gomes2021}\\
Age &9.4$\pm{1.0}$&Gyr& This work ($\S$\ref{sec:age})\\\hline
$[$Fe/H$]$ &$-$0.11$\pm$0.01 & dex &\citet{Maire2020}\\
$[$Fe/H$]$ &$-$0.09$\pm{0.04}$& dex & This work ($\S$\ref{sec:hostStarProperties})\\ \hline
log(g)&4.32$\pm$0.06&cgs &\citet{Maire2020}\\	 
log(g)&4.28$\pm{0.04}$&cgs & This work ($\S$\ref{sec:hostStarProperties})\\\hline
R.A.\ (Eq 2000; Ep 2000)&03$^h$07$^m$18.570$^s$ & &Gaia DR3\\
Dec.\ (Eq 2000; Ep 2000)&$-$13$^o$45$^\prime$42.419\arcsec & &Gaia DR3\\
Distance & 32.03$\pm$0.03 & pc & Gaia DR3\\
Proper Motion ($\mu_\alpha,\mu_\delta$)&($-$8.694, $-$240.64) &mas/yr &Gaia DR3\\
RUWE&1.0566 & &Gaia DR3\\  \hline
G & 6.814$\pm$0.003 & mag&Gaia DR3\\
H  & 5.447$\pm$0.033& mag &2MASS\\
W1 [3.4 \mum ] & 5.36$\pm$0.16&mag &WISE \\
W2 [4.6 \mum ] &5.18$\pm$0.06&mag  &WISE \\
\enddata
\end{deluxetable}

We adopted the effective temperature (\teff) and metallicity (\mh) from \citet{brewer16}, derived from a line-by-line analysis of a Keck/HIRES spectrum. Literature values from spectroscopy and Gaia color-temperature relations \citep{casagrande21} are highly consistent, with a range of  40\,K in \teff\ and 0.04\,dex in iron abundance (Table~\ref{tab:star}). We used these ranges as an estimate for uncertainties, resulting in $\teff=5747\pm40$\,K and $\mh=-0.09\pm0.04$\,dex. These uncertainties are smaller than those recommended by \citet{tayar22}, which is justified by the fact that star has properties similar to the Sun and thus suffers from smaller systematic errors. 

\begin{figure}
\begin{center}
\includegraphics[width=\linewidth]{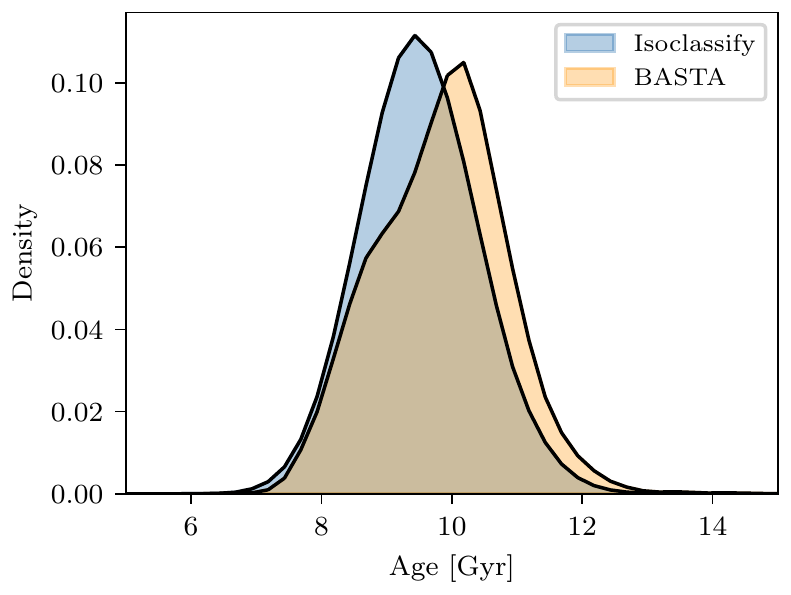}
\caption{Posterior distributions for the age of \hd\ based on isochrone modeling with isoclassify (blue) and BASTA (orange) using constraints from asteroseismology, spectroscopy and Gaia.}
\label{fig:age}
\end{center}
\end{figure}

We then combined the asteroseismic \numax\ measurement, Gaia DR3 parallax, 2MASS K-band magnitude, \teff\ and \mh\ with \texttt{isoclassify} \citep{huber17} and \texttt{BASTA} \citep{basta}, which perform Bayesian inference of stellar parameters given input observables using the stellar evolution models MIST \citep{choi16} and BASTI \citep{basti}, respectively. Importantly, \numax\  tightly constrains the surface gravity to $\logg=4.28$, which combined with the radius constraint from the Gaia parallax provides a tight constraint on stellar mass, which in turn constrains stellar age. Both tools consistently imply a mass of $\approx$\,0.95 \msun, which, given that the star has slightly evolved off the main-sequence (1.2\rsun), implies an old age. Figure \ref{fig:age} shows the age posteriors from both evolutionary models and methods. 

The final stellar parameters adopted in our study are listed in Table \ref{tab:stellar_adopted}. We adopt the self-consistent solution derived from isoclassify, but add in quadrature the difference to the BASTA results to account for systematic errors due to different model grids \citep{tayar22}.

\subsection{The Age of HD 19467}
\label{sec:age}

The age of \hd\ is important for interpreting the mass and atmospheric composition of the brown dwarf companion. As already mentioned, literature estimates have a significant spread, ranging from 5--12\,Gyr (Table~\ref{tab:stellar_adopted}). Younger, Sun-like ages come from stellar rotation \citep{Maire2020} and activity \citep{Brandt2021}, while older ages are preferred by isochrone fitting \citep{Wood2019, Maire2020}. The rotation-age is based on the detection of photometric period of $\approx$\,29 days with an amplitude of $\approx$\,0.5\% from ground-based ASAS data \citep{Maire2020}. The high-precision of TESS light curve in Figure \ref{fig:lc} rules out rotational modulation at the level of 0.5\% over 25 day timescale suggested by the ASAS data, which implies that the rotation period for \hd\ is undetermined. This is consistent with results from the Kepler Mission, which demonstrated that typical rotational amplitudes in mature Sun-like stars are on the order of a few hundred ppm \citep{mcquillan14,santos21} and thus are generally not detectable using ground-based photometry. Chromospheric activity-based ages also become more challenging for stars with Sun-like and older ages due the flattening of the age-activity relation, making age constraints sensitive to small changes in $R'_{HK}$ measurements. While some literature values for $R'_{HK}$ favor near solar values (and thus ages) for \hd, others are consistent with older, isochrone-based ages \citep[e.g.\ $R'_{HK}=5.1$ and $8.8\pm0.3$\,Gyr,][]{lorenzo18}.

The asteroseismic detection from TESS supports an older age for \hd. While the low S/N precludes a direct age from a measurement of individual oscillation frequencies \citep[e.g.][]{mathur12,metcalfe14,silva17}, the \numax\ measurement precisely constrains \logg\ and thus stellar mass independent of stellar evolutionary models. With a mass similar to solar (\massstar \msun), \hd\ must have an age significantly older than the Sun to reach a radius of 
1.2\rsun \footnote{This analogy is only slightly affected by the sub-stellar metallicity of \hd; a solar-mass star with \fehstar\ has an age of $\approx$\,3.6\,Gyr at solar radius.}. 
As discussed by \citet{Maire2020}, an age of \agestar\,Gyr is compatible with the slight enhancements in alpha elements and sub-solar metallicity, placing the star in the transition region between chemical ``thin-disk'' and ``thick-disk'' stars. Overall, we conclude that \hd\ is an $\approx$4-5\,Gyr older analog to our Sun and adopt an age of 9.4$\pm$1.0 Gyr (Table~\ref{tab:stellar_adopted}).

\begin{table}
\begin{center}
\caption{Adopted Stellar Parameters for \hd A
\label{tab:stellar_adopted}}
\renewcommand{\tabcolsep}{0mm}
\begin{tabular}{l c}
\tableline\tableline
\noalign{\smallskip}
Effective temperature, \teff\, (K) & \teffstar \\
Metallicity, \mh\ (dex) & \fehstar \\
\hline
Luminosity, $L$ (\lsun) & \lumstar \\
Stellar radius, \rstar\ (\rsun)& \radstar \\
Stellar mass, \mstar\ (\msun)& \massstar \\
Stellar density, \rhostar\ (cgs)& \denstar \\
Surface gravity, \logg\ (cgs) & \loggstar \\
Age, $t$ (Gyr) & \agestar \\
\noalign{\smallskip}
\tableline\tableline
\end{tabular}
\end{center}
\tablecomments{\teff\ and \mh\ are adopted from \citet{brewer16}, with uncertainties accounting for the spread in literature results. All other properties are derived from the combination of constraints from asteroseismology, spectroscopy and Gaia (see \S\ref{sec:hostStar}).} 
\end{table}

\section{NIRCam Data Reduction and Post Processing}\label{sec:NIRCAMresults}

We use the processed images retrieved from the Mikulski Archive for Space Telescopes (MAST)\footnote{https://mast.stsci.edu/} that have been corrected for bad pixels, flat-fielding, and background subtraction with the \texttt{jwst} pipeline. The product we use are the \texttt{calints} files which  result from \textit{Stage 2} of the pipeline and have been through a photometric calibration. The data were processed with calibrations software version 1.5.3 and calibration reference data context \texttt{jwst\_0943.pmap}. In addition to these data, we take advantage of wavefront information provided by Optical Path Difference (OPD) maps taken by the NIRCam wavefront sensing team\footnote{https://webbpsf.readthedocs.io/en/latest/available\_opds.html} to 
generate a NIRCam PSF model close in time to our science observations.
For synthetic PSFs, we utilize the OPD from 2022-08-11, \textit{R2022081102-NRCA3\_FP1-1.fits}, the closest in time preceding our observations.

\subsection{PSF Subtraction \label{sec:pca}}

\begin{figure*}
\includegraphics[width=0.95\textwidth]{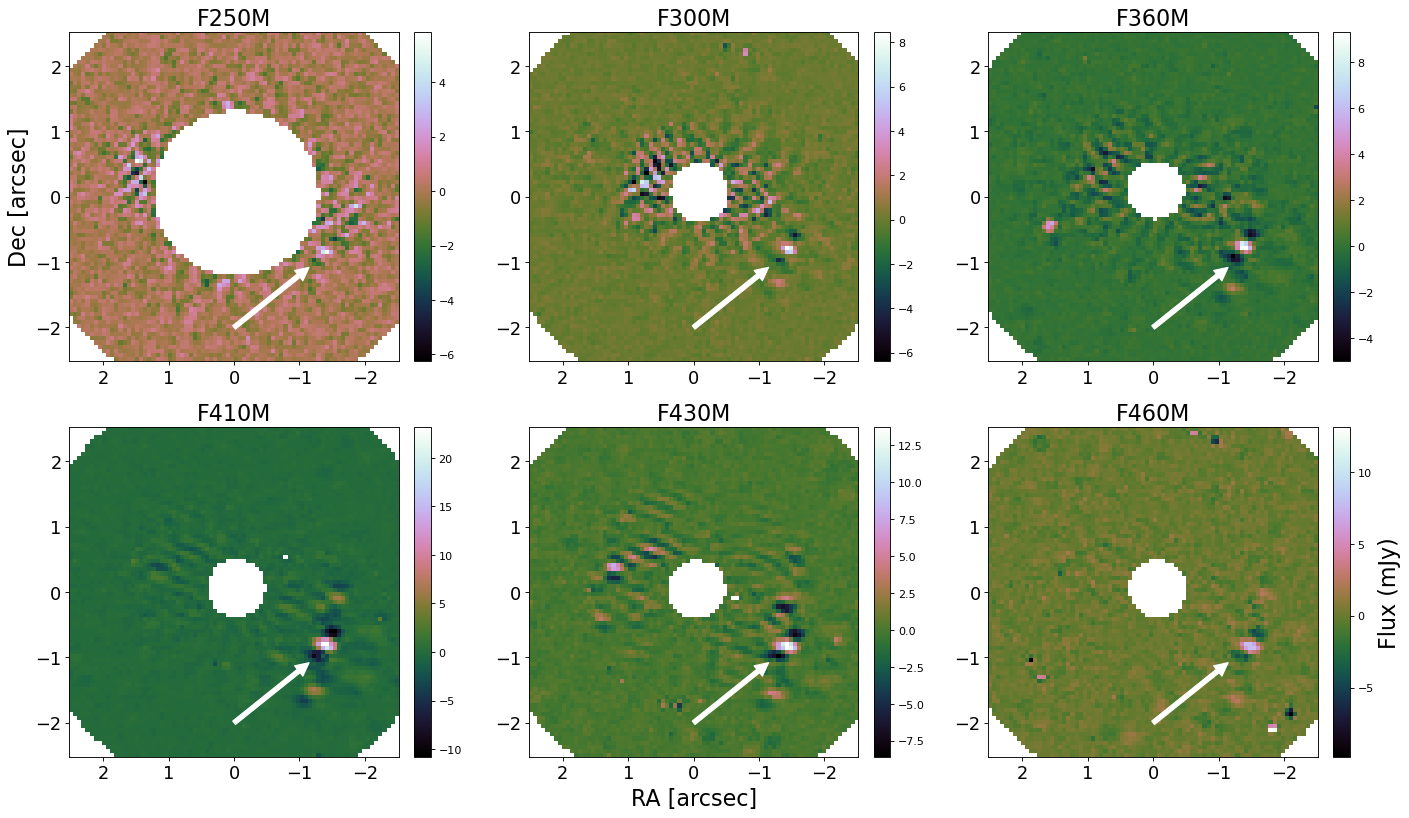} 
\caption{Post-processed images of \hd\ in the six NIRCam filters observed in this program, rotated so that North is up. Images were reduced using the pyKLIP algorithm as described in the text. Arrows indicate the detected companion. \label{fig:NIRcamPCA}}
\end{figure*}

We apply principal component analysis (PCA) \citep[e.g.][]{Lafreniere2007,Amara2012} via Karhunen Lo\'eve Image Projection \citep[KLIP;][]{Soummer2012}, to subtract the residual stellar intensity from the science frames using the images taken in two roll angles for angular diversity. We perform PSF subtraction using the open source Python package \texttt{pyKLIP} \citep{Wang2015} using Angular Differential Imaging (ADI) and Reference Differential Imaging (RDI) using a synthetic reference PSF, as described below. The results of the PCA reduction for all filters is displayed in Figure~\ref{fig:NIRcamPCA}.

\begin{figure*}
\begin{center}
\includegraphics[width=0.8\linewidth]{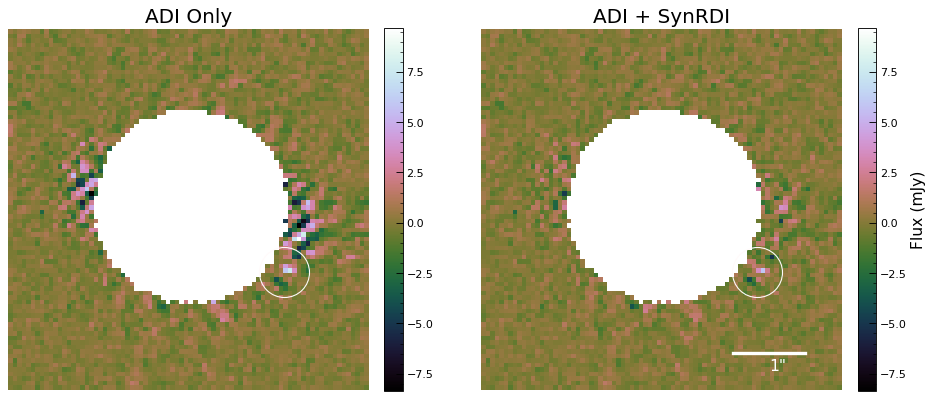} 
\includegraphics[width=0.5\linewidth]{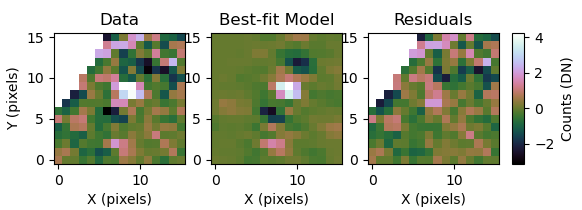}
\caption{\textbf{Top:} Comparison between using only ADI for the PCA reduction (\textit{left}), and using RDI with synthetic PSFs generated with \texttt{WebbPSF} (\textit{right}) for the the F250M data. The addition of RDI reduces the speckle noise in the PSF-subtracted images. The data are oriented so that North is up. 
\textbf{Bottom:} The forward model compared with the PSF-subtracted data for F250M, using the sythetic PSFs as reference.}
\label{fig:f250m_adi_rdi}
\end{center}
\end{figure*}

For all filters except filter F250M, the data contained in the two rolls suffices to obtain an unambiguous detection of the companion. For the F250M case, although the companion's signal is visible using only ADI, we resorted to using RDI with a set of synthetic PSFs in order to confirm the signal is indeed from the companion and not due to residual speckles. In Figure~\ref{fig:f250m_adi_rdi} we show the comparison, for the F250M filter, between only using the roll frames for the PCA reduction (ADI), and assisting the PCA reduction with a set of synthetic PSFs (ADI+SynRDI).

The grid of synthetic stellar PSFs is generated using \texttt{WebbPSF} \citep{Perrin2014} and tools from \texttt{webbpsf\_ext}\footnote{https://github.com/JarronL/webbpsf\_ext} at offset locations with respect to the coronagraph focal plane mask. We generate simulated PSFs in different sets of 9-point grid pattern at even spacings. We simulate spacings of 2.5, 7, 15, 25, 40 mas, in addition to a set of rotations of the coronagraphic-PSF with respect to the detector of 0.1, 0.3, 0.5 degrees. 
This aims to emulate the speckles present in the data frames, and assists the PCA reduction with the diversity in speckle structure needed to perform a more optimal reference subtraction. 

As mentioned above, for filters F300M, F360M, F410M, F430M, and F460M, ADI suffices for a clear detection. As a second step we use RDI with the synthetic PSFs to further subtract the unwanted starlight. This is motivated by the fact that the companion PSF's northern lobe falls near the diffraction speckles caused by the bar coronagraph. The number of Karhunen Lo\'eve (KL) modes determines how much of the synthetic PSFs are used for the subtraction. Since these have been generated with arbitrary offsets, there is a risk of subtracting the light from the secondary. We use 15 KL modes, which  minimizes over-subtraction  and  clears out slightly more of the residual starlight around the northern lobe. This was done by visual inspection; a more in-depth analysis on how to optimally use synthetic PSFs will be explored in the future.

\subsection{Photometry \label{sec:photometry}}

To accurately extract the flux and position of \hdb, we account for over-subtraction effects on the PSF that arise during the reduction process (described in Section~\ref{sec:pca}) with a forward model based on the method described by \cite{Pueyo2016}. We make use of its implementation on \texttt{pyKLIP} \citep{Wang2015}. The companion PSF is modeled using \texttt{WebbPSF} \citep{Perrin2014} for each filter and accounting for its position with respect to the bar focal plane mask. An accurate position of the simulated PSF is particularly important in the case of the shorter wavelength filters: poorer spatial sampling of the pixels compared to the diffraction limit at smaller wavelengths (2.5\um is sub-Nyquist) results in an acute sensitivity of the PSF structure as seen in the detector. An accurate positioning for the case of F250M was done by trial and error simulating a grid of PSF offsets and selecting the best fit by least square difference between the simulation and the coadded science frames. The model PSF are simulated using the OPD map closest in time and prior to the observations.

\begin{figure}
\begin{center}
\includegraphics[width=\linewidth]{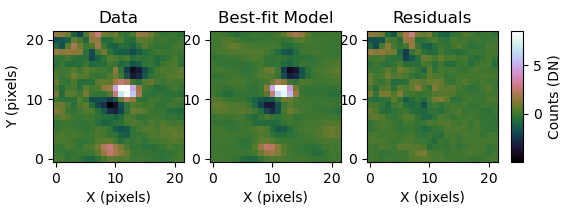}
\caption{The best fit model to the PSF-subtracted signal used to measure the photometry and astrometry of \hdb\ in filter F360M, where the highest SNR was achieved. The residuals show the companion is fit well by the forward model.}
\label{fig:photometry_astrometry_extraction}
\end{center}
\end{figure}

The flux and position of the companion are extracted with \texttt{pyKLIP}. We fit a model of its photometry and astrometry to the reduced data using an MCMC approach (\texttt{emcee;} \cite{Foreman-Mackey2013}). Figure~\ref{fig:photometry_astrometry_extraction} shows the PSF model fit to the reduced data for filter F360M, the filter in which we obtain the highest SNR. Appendix \ref{sec:fmstamps} contains the full gallery of forward model comparisons with the PSF-subtracted images in each band.

The flux calibration of the signal is determined based on the \texttt{jwst} stage 2 pipeline photometric calibration. We apply a flux correction to the photometry based on measured attenuation factor of $0.92$ of the Bar mask Lyot stop at $\sim1.6$\arcsec. We fit a G3V stellar photosphere model to 1-5 \um photometry from 2MASS \citep{Cutri2003} and WISE \citep{Cutri2012}. We also find a $\sim2\%$ error in fitting the stellar model to the IR measurements, and apply this error to contrast reported.
Table \ref{tab:stellar_flux} shows the estimated stellar flux. 
Comparison of the calibrated flux measured from the acquisition and astrometric confirmation images, both taken through the neutral density square, produced from the stage 2 pipeline is consistent with the estimated stellar spectrum within $\sim10\%$.
We therefore apply a $10\%$ uncertainty to reported absolute photometry of \hdb\ in this section.

\begin{table}
\begin{center}
\caption{Adopted Photometry For \hd\ A
\label{tab:stellar_flux}}
\renewcommand{\tabcolsep}{3mm}
\begin{tabular}{c c}
\tableline\tableline
\noalign{\smallskip}
Filter & Flux (Jy) \\ \hline
F250M flux (Jy) & 3.51$\pm$0.07 \\  
F300M flux (Jy) & 2.63$\pm$0.05 \\  
F335M flux (Jy) & 2.10$\pm$0.04 \\ 
F360M flux (Jy) & 1.82$\pm$0.04 \\ 
F410M flux (Jy) & 1.49$\pm$0.03 \\ 
F430M flux (Jy) & 1.36$\pm$0.03 \\ 
F460M flux (Jy) & 1.12$\pm$0.02 \\ 
\noalign{\smallskip}
\tableline\tableline
\end{tabular}
\end{center}
\tablecomments{Predicted fluxes in JWST wavebands are based on BOSZ stellar models \citep{bohlin17}.}
\end{table}

\begin{figure}
\includegraphics[width=0.45\textwidth]{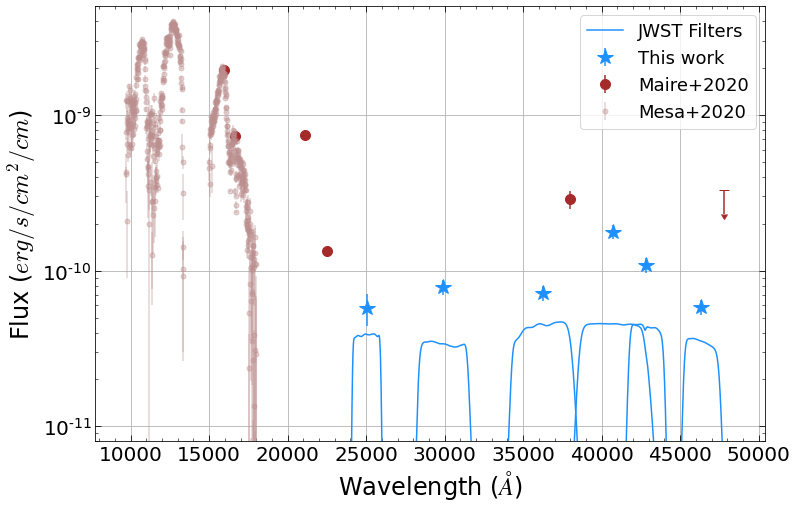} 
\caption{New NIRCam photometry (blue stars) compared with recent ground-based measurements from VLT-SPHERE and VLT-NACO. 
\label{fig:photresults}}
\end{figure}

Figure \ref{fig:photresults} shows the measured photometry in each NIRCam band alongside recent measurements and limits from the ground \citep{mesa2020, Maire2020}. The F460M flux is consistent with the M band upper limit obtained with VLT NaCo \cite{Maire2020}, however there appears to be some tension with the NaCo L' flux compared with F360M and F410M photometry measurements. \citet{Carter2022} also noted a discrepancy in measurements from NaCo $L'$ and JWST NIRCam photometry. The difference in passband on a steeply rising part of the spectrum, possible water vapor effects, as well as calibration uncertainties may account for this discrepancy. Continued refinement of JWST photometric calibrations will help identify any biases in photometry. For this study, we do not incorporate NaCo photometry into the analysis, but rather present the measurement comparison for future investigation. 

Table \ref{tab:NIRCamresults} shows our measured photometry and relative astrometry for \hdb\ (see next section).

\begin{deluxetable*}{l|ll|ll}
\tabletypesize{\scriptsize}
\tablewidth{0pt}
\tablecaption{NIRCam measurements of \hdb\ \label{tab:NIRCamresults}}
\tablehead{\colhead{}& \colhead{Separation}& \colhead{Pos.\ Angle}&\colhead{$\Delta$mag}&\colhead{Flux}\\
\colhead{Filter}  &  \colhead{(\arcsec)} & \colhead{(deg)}&\colhead{(mag)}&\colhead{($\mu$Jy)}}
\startdata
F250M & 1.597$\pm$0.010 & 236.9$\pm$0.14 & 13.67$\pm$0.271  & 11.96$\pm$2.75 \\ 
F300M & 1.611$\pm$0.002 & 237.2$\pm$0.04 & 12.62$\pm$0.122 & 23.50$\pm$2.52 \\ 
F360M & 1.610$\pm$0.003 & 236.9$\pm$0.07 & 11.91$\pm$0.127  & 31.41$\pm$3.58 \\ 
F410M & 1.604$\pm$0.001 & 236.8$\pm$0.04 & 10.45$\pm$0.116  & 98.74$\pm$10.20 \\ 
F430M & 1.609$\pm$0.003 & 236.6$\pm$0.07 & 10.78$\pm$0.124  & 66.32$\pm$7.35 \\ 
F460M & 1.609$\pm$0.003 & 236.9$\pm$0.07 & 11.07$\pm$0.121  & 41.78$\pm$4.64 \\ 
\hline \hline
\enddata
\tablecomments{NIRCam astrometry and photometry from 2022-Aug-12}
\tablecomments{The astrometric precision for each filter is based solely on the positional uncertainty relative to the center of the coronagraph mask.  The combined astrometry includes an additional term to account for uncertainty in the stellar position behind the mask (7 mas in each direction).}
\end{deluxetable*}

\subsection{Relative Astrometry}

\begin{figure}
    \centering
    \includegraphics[width=0.45\textwidth]{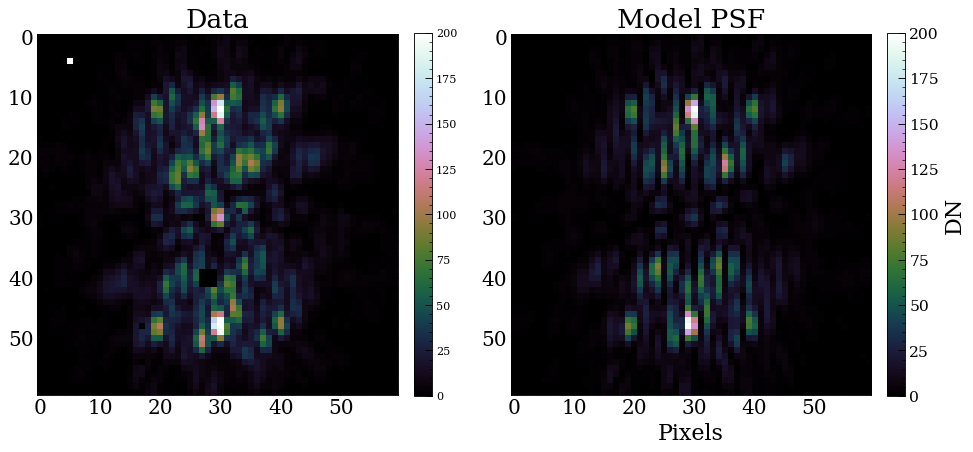}
    \caption{\textbf{Left:} An example of the raw MASKLWB coronagraph data for an image in the F360M filter. \textbf{Right:} The best fit model PSF simulated using the most recent preceding OPD map, used to measure the centroid of the star.}
    \label{fig:centroiding}

\end{figure}

A major challenge of obtaining relative astrometry of a companion in coronagraphic imaging is that the primary star is occulted by the focal plane mask. Knowledge of the wavefront from published OPD maps, and a highly structured PSF enable a forward model based cross-correlation with the data to fit for the centroid of the star behind the mask. We perform a cross-correlation of model PSFs with the data using the \texttt{chi2\_shift} in the \texttt{image-registration} Python package\footnote{https://image-registration.readthedocs.io/} to measure the best fit position of the star behind the mask (Figure \ref{fig:centroiding}).
We obtain a centroiding error $\sim$7 $\mathrm{mas}$, consistent with the measured sensitivity in \citet{Carter2022}.

\begin{figure}
    \centering
    \includegraphics[width=0.45\textwidth]{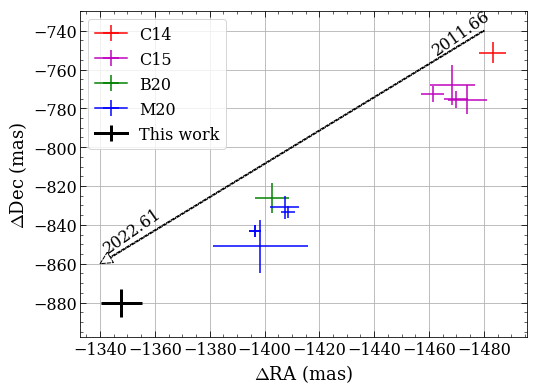}
    \caption{Astrometric position of \hdb\ relative to its parent star, compared with previous measurements.}
    \label{fig:comboastrometry}
\end{figure}

The companion position is recovered 
with the joint astrometry and photometry model fit to the reduced data as described in Section~\ref{sec:photometry}. The model fit errors provide the uncertainty in the relative position to the measured star position on the detector. We add the star position uncertainty to the reported errors (Table \ref{tab:NIRCamresults}). Figure~\ref{fig:comboastrometry} shows the new astrometric measurement compared to previous relative astrometry measurements of \hdb\ \citep{Crepp2014, Crepp2015, Bowler2020, Maire2020}.

\subsection{Performance and Sensitivity}
\begin{figure*}
\centering
\includegraphics[width=0.79\textwidth]{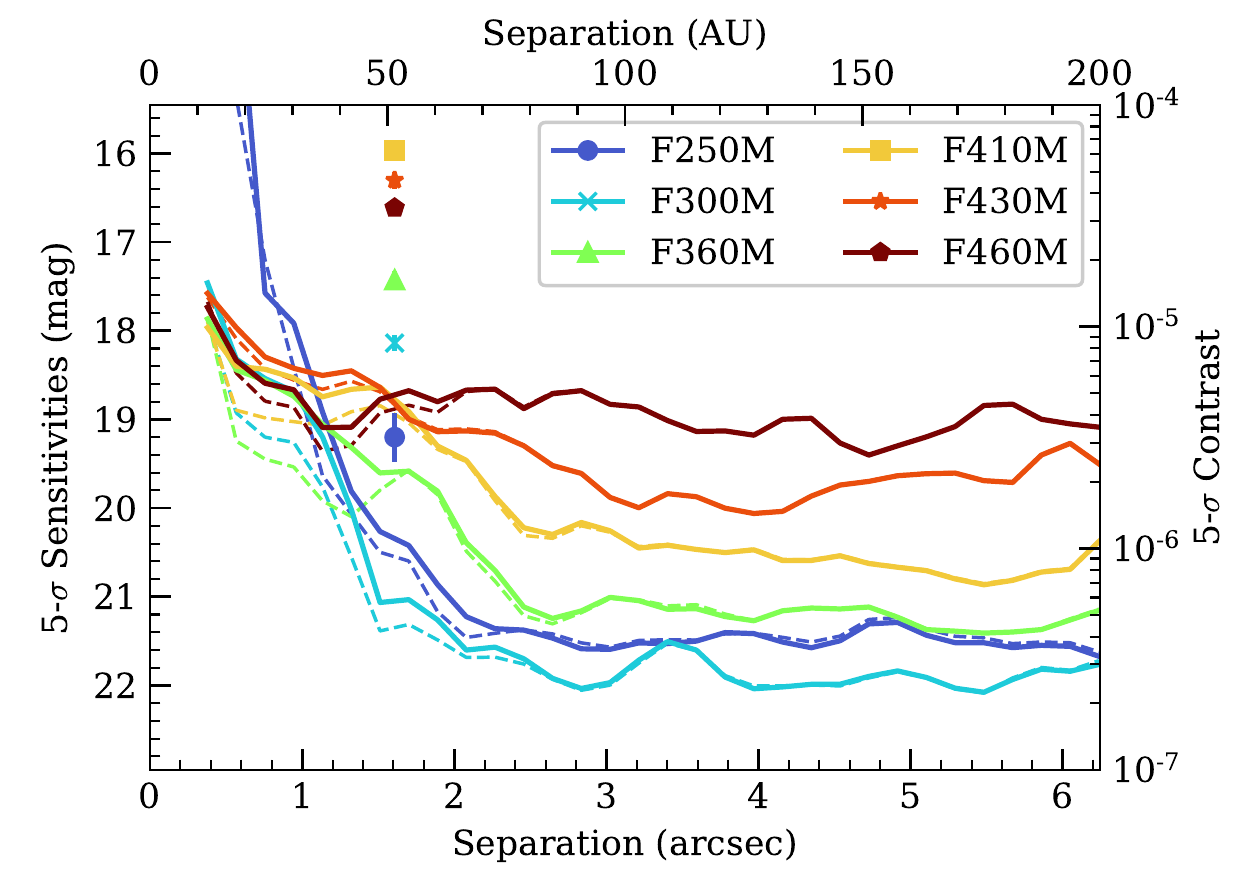} \caption{Contrast curves for all filters. Solid lines indicate ADI, and dashed lines indicate ADI and RDI using synthetic PSFs. Data points indicate the \hdb\ detections. The use of synthetic PSFs provides the diversity necessary to obtain enhanced contrast at small angular separations. 
\label{fig:contrast_curves}}
\end{figure*}

In Figure~\ref{fig:contrast_curves} we show the contrast curves for the reduced images after PSF subraction. The contrast is measured with \texttt{PyKLIP} by computing the noise in an azimuthal annulus at each separation, using a Gaussian cross correlation to remove high frequency noise. The flux normalization to obtain these contrast numbers was computed as explained in Section~\ref{sec:photometry}, by using a best fit model of the stellar spectrum to calibrate contrast. The contrast curves are corrected for algorithmic throughput, i.e.\ the throughput loss due to the PSF subtraction, and for small sample statistics \citep{Mawet2014}.

\begin{figure*}
\centering
\includegraphics[width=0.49\textwidth]{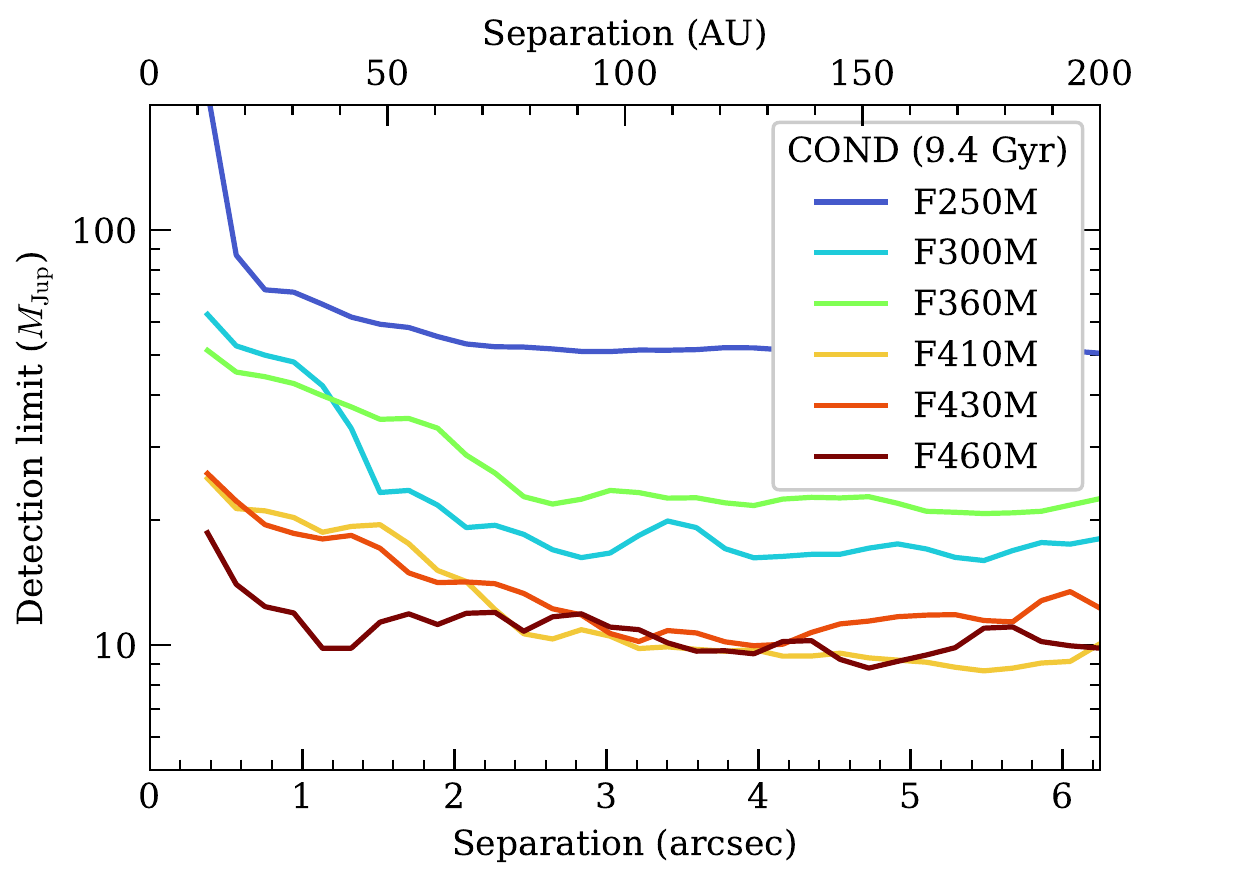} 
\includegraphics[width=0.49\textwidth]{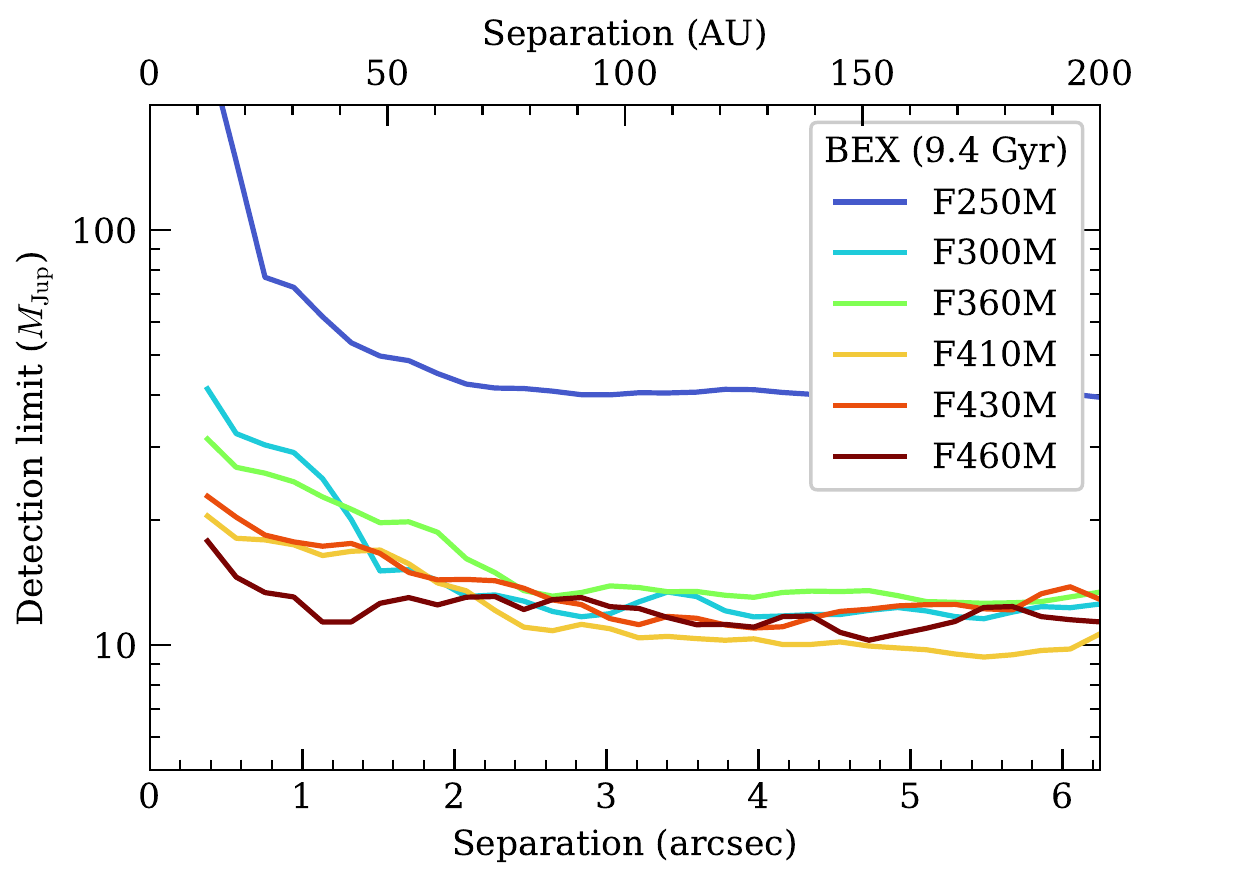} 
\includegraphics[width=0.49\textwidth]{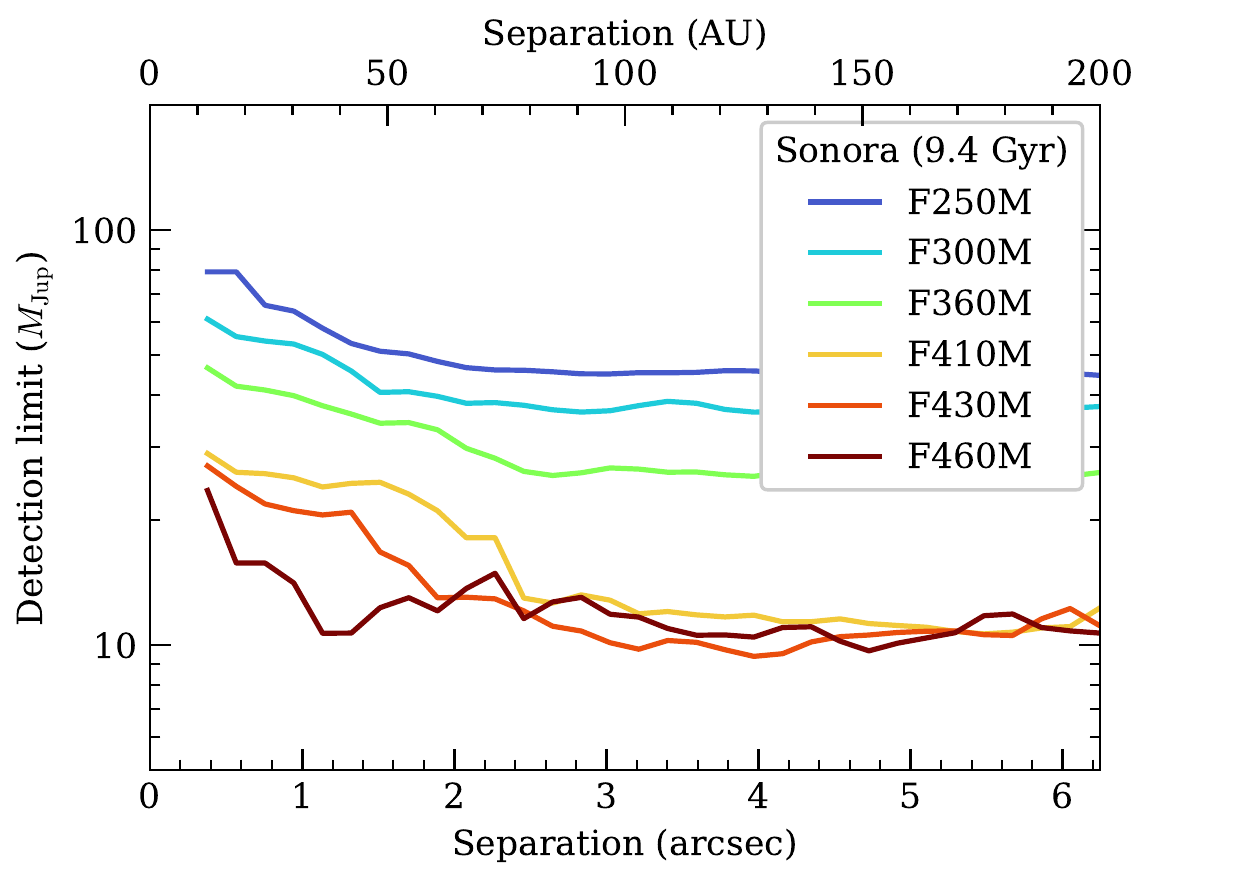}
\includegraphics[width=0.49\textwidth]{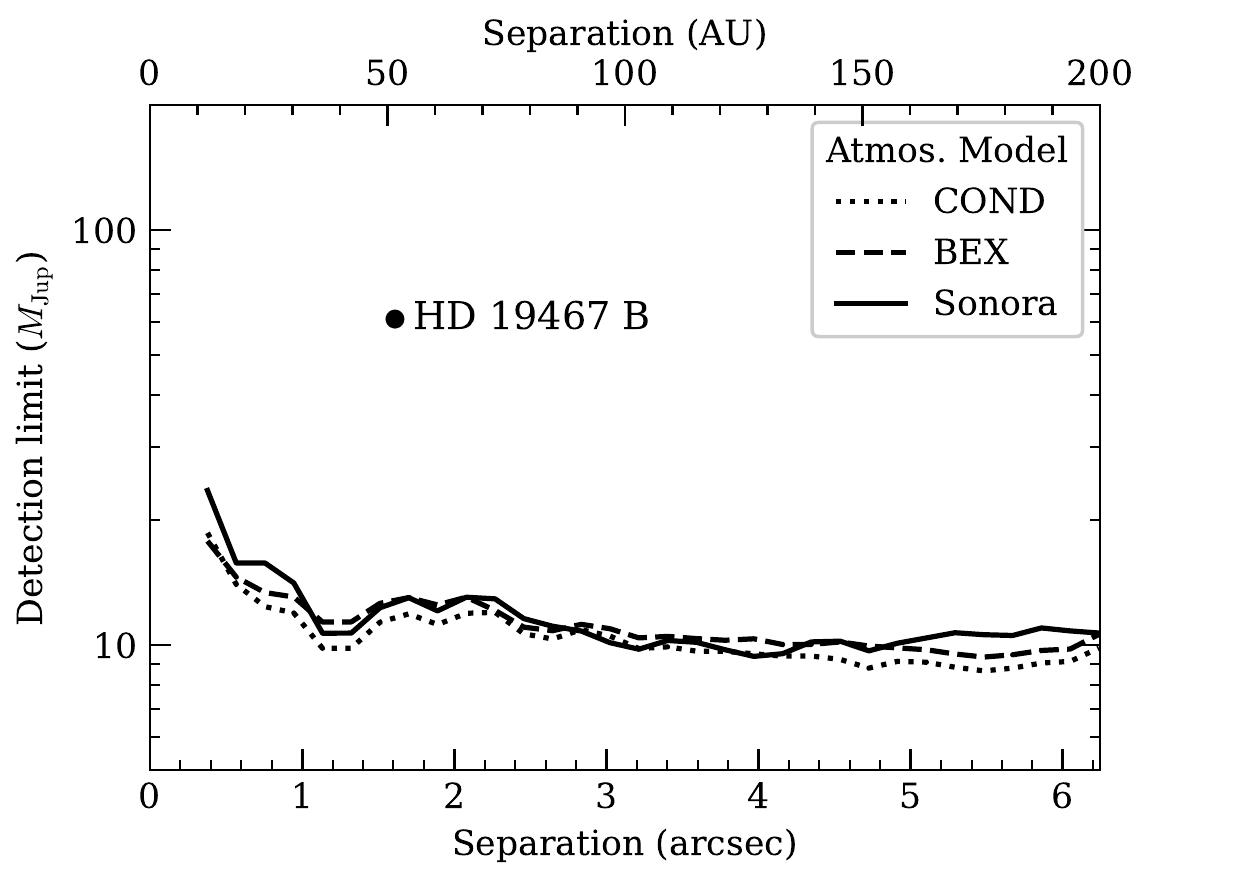}
\caption{Detection limits ($5-\sigma$) for each filter, in terms of companion mass.
The first three panels show limits for three different atmospheric evolution models (COND, Bex, and Sonora), while the last panel compares
the overall detection limit for each of the models. 
A mass estimate from atmospheric model fitting (Section \ref{sec:atmosphereFitting}) is shown as a point in the lower right figure (61~\mj\ at 1.61\arcsec).
\label{fig:mass_limits}}
\end{figure*}

Figure \ref{fig:mass_limits} translates our detection limits from flux/contrast sensitivities to limits on companion mass.
Three different brown dwarf evolution models are considered --
Ames-COND \citep{Baraffe2003},
BEX-HELIOS \citep{Linder2019},
and Sonora-Bobcat \citep{marley2021}.
In each case, we assume solar metallicity.

While the shortest wavelength observations achieve the best contrast (in particular F300M), the longer wavelengths are better at detecting lower mass companions.  
We find an overall detection limit of $\sim$10 \mj.
This limit is much higher than the sub-Jupiter levels that JWST/NIRCam can obtain for young systems \citep[e.g.][]{Carter2022}, but even for this very old system brown dwarfs are easily detectable outside of $\sim$0.4\arcsec\ $\simeq$ 10 AU.
While the detection limit is relatively independent of model, it does depend significantly on the age of the brown dwarf (the system age is discussed in \S\ref{sec:age}).

Despite a lack of reference star observations, we are able to recover the signal of \hdb\ with two roll angles and achieve contrasts $\sim$10$^{-5}$ at 1--2 arcsec. Regular OPD measurements enable the use of synthetic PSFs that can aid PSF subtraction by generating a set of reference PSFs to capture speckle structure. This suggests that bright companions could be observed without reference stars, significantly reducing the time spent on the observation. Future work will investigate the difference between reducing data with and without reference star observations. Future observations with a better defined position for the LWB coronagraph should also provide better contrast close-in.

\section{Orbit of HD 19467 B}
\label{sec:orbitFitting}

Previous studies estimate the mass of \hdb\ from 51 to 86\mj\ through both model-based estimates and orbital analyses \citep{Crepp2014, Maire2020, Brandt2021}. We analyze new radial velocities and provide an updated dynamical mass estimate including our new relative astrometry and additional RV measurements. 

\begin{figure}%
 \centering
\includegraphics[width=0.45\textwidth]{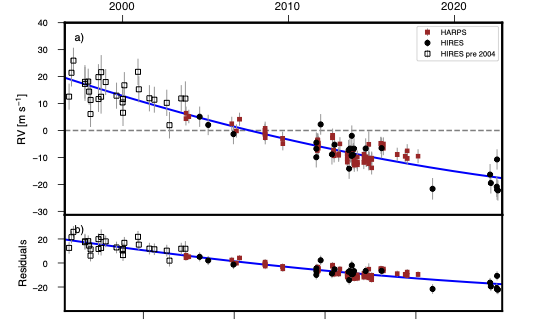} 
\caption{Best fit single companion Keplerian orbital model for \hd\ to radial velocities measured for \hd\ by HIRES by the California Legacy Survey \citep{Rosenthal2021}, new HIRES observations, and HARPS \citep{trifonov2020}. The fit slightly favors a curvature term.  \label{fig:RVdata}}
\end{figure}

First, we fit the new and previously measured RVs from HIRES and HARPS \citep{trifonov2020} using the \texttt{RadVel}\footnote{https://radvel.readthedocs.io/en/latest/} software \citep{fulton2018}. With the addition of the new data, we measure a linear slope term of $\dot{\gamma}=-0.00412\pm0.00027$ m s$^{-1}$ d$^{-1}$ with strong significance. We attempt to fit for curvature and tentatively detect a curvature term of $\ddot{\gamma}=1.7^{+0.81}_{-0.78} \times 10^{-7}$ m s$^{-1}$ d$^{-2}$ at 2.1$\sigma$. Model comparison using $\Delta BIC$ and $\Delta AIC$ (Aikike Information Criterion, \cite{aic}) show a nearly indistinguishable model fit to a trend-only and trend plus curvature model. A detection of curvature can place strong constraints on the companion orbit, especially for higher eccentricity systems. Figure \ref{fig:RVdata} shows the RV data plotted over the maximum likelihood model. 
Appendix \ref{sec:RVfit} contains a more detailed description of the fit comparison and the new radial velocities used in the analysis.

For the full orbital analysis we include all available RV measurements from HARPS \citep{trifonov2020} and HIRES (HIRES data including new measurements tabulated in Appendix \ref{sec:RVapp}), relative astrometry  \citep{Crepp2014, Crepp2015, Bowler2020, Maire2020} (listed in Table \ref{tab:astrometry}), and absolute astrometry from Hipparcos and Gaia as described in
\citet{Brandt2021}, which takes advantage of proper motion anomalies between Hipparcos, Gaia EDR3 and the Hipparcos-Gaia long-term trend.
We utilize the cross-calibrated catalog of Hipparcos-Gaia accelerations presented in \cite{BrandtT2021}.

\begin{deluxetable}{lcllll}
\tabletypesize{\scriptsize}
\tablewidth{0pt}
\tablecaption{Imaging Astrometry\label{tab:astrometry}}
\tablehead{
\colhead{epoch$-2450000$} & \colhead{Filter}& \colhead{$\rho$ (mas)} & \colhead{$\rho_{err}$} & \colhead{PA (deg)}& \colhead{PA$_{err}$} }
\startdata
\multicolumn{6}{l}{\textit{Astrometry from \citet{Crepp2014}}}\\
5804.1 & K' & 1662.7 & 4.9 & 243.14 & 0.19 \\
5933.8 & H &  1665.7 & 7.0 & 242.25 & 0.26 \\
5933.8 & K' & 1657.3 & 7.2 & 242.39 & 0.38 \\
6166.1 & K' & 1661.8 & 4.4 & 242.19 & 0.15 \\
6205.0 & Ks & 1653.1 & 4.1 & 242.13 & 0.14 \\
\multicolumn{6}{l}{\textit{Astrometry from \citet{Maire2020}}}\\
8032.3 & L' &  1637    & 19   & 238.68 & 0.47  \\
8061.2 & K1 &  1636.7  & 1.8  & 239.39 & 0.13  \\
8061.2 & K2 &  1634.4  & 5.0  & 239.44 & 0.21  \\
8409.3 & H2 &  1631.4  & 1.6  & 238.88 & 0.12 \\
8409.3 & H3 &  1631.4  & 1.6  & 238.88 & 0.12  \\
\multicolumn{6}{l}{\textit{New Astrometry (this work; see Table \ref{tab:NIRCamresults}) }}\\
9803.9 & 2.5--4.6\mum & 1607.6 & 7 & 236.84 & 0.25 \\
\hline
\enddata
\end{deluxetable}

We use \texttt{orvara} \citep{BrandtOrvara} to fit orbits to the radial velocities, absolute astrometry, and relative astrometry.  \texttt{orvara} is an orbit fitting code that uses \texttt{ptemcee}, a parallel tempered MCMC scheme \citep{Foreman-Mackey2013, Vousden2016}. Following the orbital analysis in \citet{Brandt2021}, we apply a geometric prior to inclination and log-flat priors to semi-major axis and companion mass. We apply uniform priors to remaining orbital elements. log-flat priors are applied to RV jitter. We adopt the mass of $M_*=$~0.96~$\pm0.02~\msun$ based on the analysis in \S\ref{sec:hostStar} using asteroseismology, spectroscopy, and Gaia data. 

\begin{figure}
    \centering
    \includegraphics[width=0.4\textwidth]{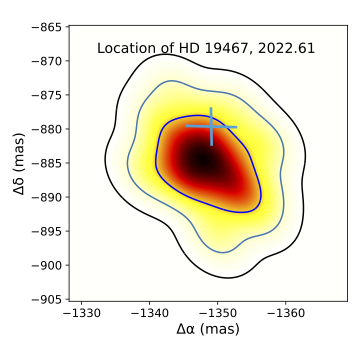}
    \caption{Orvara prediction of relative astrometry at the 2022-08-12 epoch. The measured astrometry is consistent with the prediction. }
    \label{fig:prediction}
\end{figure}

We first predict the position of \hdb\ in the current epoch leaving out the new relative astrometry measured with NIRCam, but including all other data. Figure \ref{fig:prediction} shows that our measurement is consistent with the prediction of the best fit orbits using previous measurements. 

Next, we fit for orbital parameters including our new relative astrometry measurement from NIRCam. Table \ref{tab:orvara} summarizes the orbit fit results. We infer a mass of $81^{+14}_{-12}$\mj. Our mass estimate for \hdb\ is within 1-$\sigma$ of the prior estimates from in \citet{Brandt2021} ($65.4^{+5.9}_{-4.6}$\mj), and \citet{Maire2020} ($74^{+12}_{-9}$\mj). We infer an eccentricity of $0.416^{+0.092}_{-0.07}$, which is consistent with recent measurements in \citet{Brandt2021}, $0.54\pm0.11$, \citet{Maire2020}, $0.56\pm0.09$, and \citet{Bowler2020}, $0.39^{+0.26}_{-0.18}$. We infer a period of $386^{+220}_{-108}$~yr, which is consistent with prior orbital analyses \citep{Bowler2020, Maire2020, Brandt2021}. The tentative evidence for curvature from the new radial velocities may indicate that the orbit is close to periastron passage. Given the high eccentricity, it could be a critical time to monitor this system.

\begin{figure}[b!]%
\includegraphics[width=0.45\textwidth]{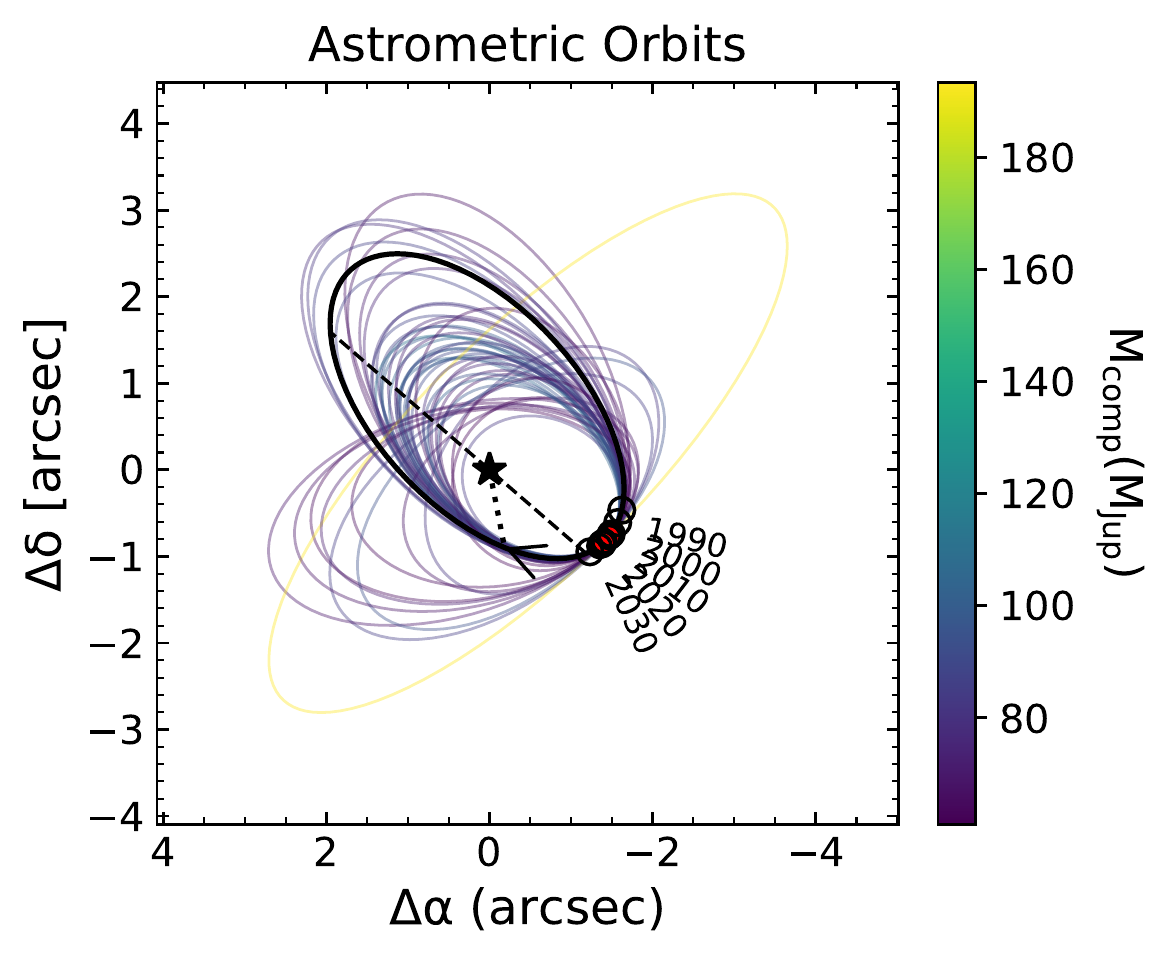} 
\caption{A selection of orbits from the MCMC posteriors. \label{fig:orbitplot}} 
\end{figure}

\begin{figure*}%
\begin{centering}
\includegraphics[width=0.9\textwidth]{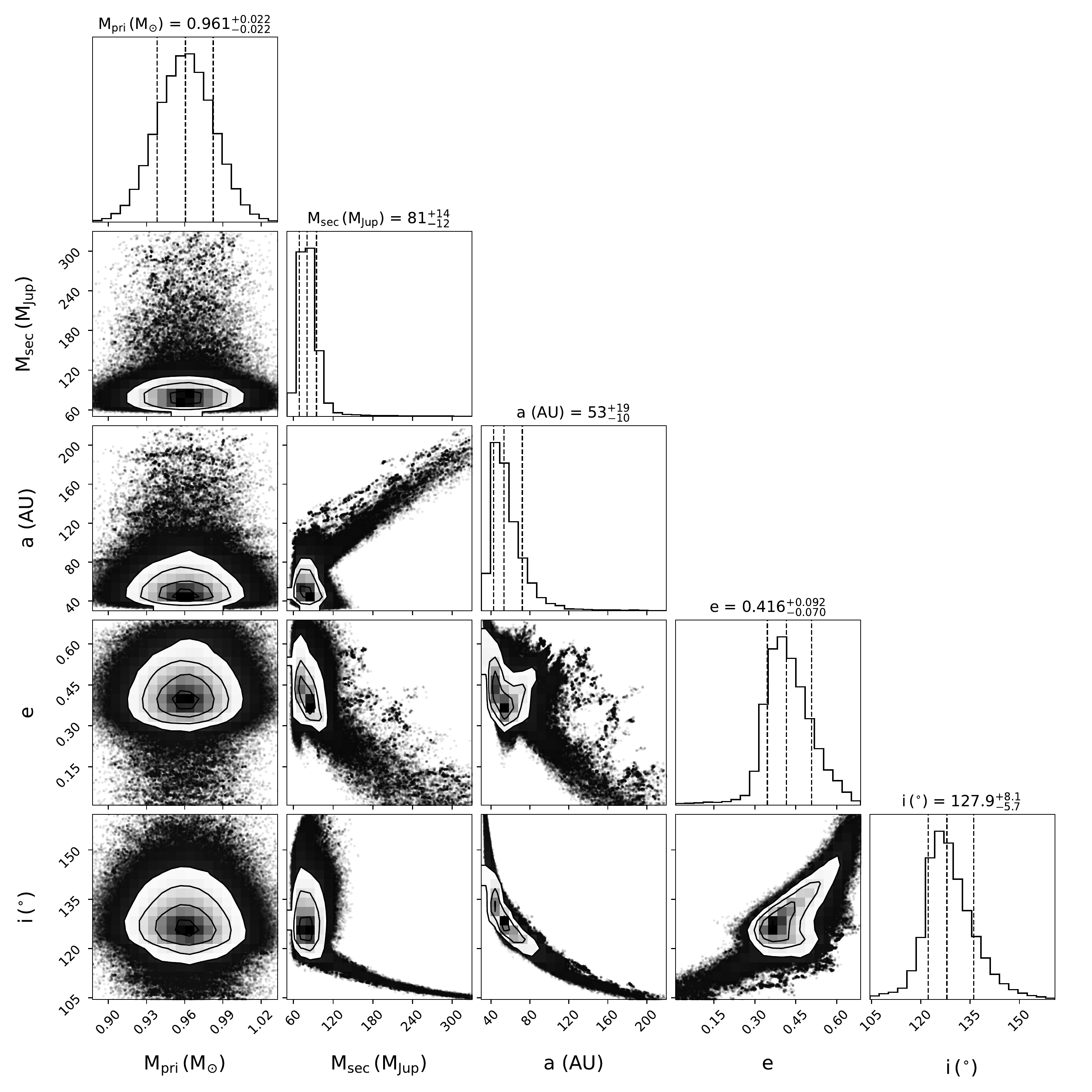} 
\caption{Corner plot showing MCMC posteriors for select orbital parameters. \label{fig:orbitcorner}}
\end{centering}
\end{figure*}

Figure \ref{fig:orbitplot} shows a selection of orbits from MCMC posteriors overlaid with the relative astrometry used in the fit, and Figure \ref{fig:orbitcorner} displays a corner plot of the MCMC posteriors for orbital parameters.

\begin{deluxetable}{lll}
\tabletypesize{\scriptsize}
\tablewidth{0pt}
\tablecaption{Orbit fit results\label{tab:orvara}
}
\tablehead{
\colhead{Parameter}& \colhead{Units} & \colhead{Value}}
\startdata
Jitter & m/s  & ${3.17}_{-0.23}^{+0.25}$\\
$M_{pri}$ & $M_*$ & ${0.961}_{-0.022}^{+0.022}$ \\
$M_{sec}$ & \mj  &       ${81}_{-12}^{+14}$ \\
Semi-major Axis & AU & ${53}_{-10}^{+19}$ \\
$\sqrt{e}\sin\omega$  & & ${-0.586}_{-0.061}^{+0.068}$ \\
$\sqrt{e}\cos\omega$ & & ${0.08}_{-0.32}^{+0.29}$ \\
Inclination & deg &   ${127.9}_{-5.7}^{+8.1}$\\
Ascending Node & deg & ${48.9}_{-7.3}^{+240}$ \\
Mean Longitude & deg &    ${192.7}_{-123}^{+7.7}$ \\
Parallax & mas & ${31.226}_{-0.037}^{+0.037}$ \\
Period & year & ${382}_{-108}^{+220}$ \\
Argument of Periastron & deg &  ${278}_{-29}^{+27}$ \\
Eccentricity & &  ${0.416}_{-0.070}^{+0.092}$ \\
Semi-major Axis & mas & ${1664}_{-328}^{+598}$ \\
$T_0$ & JD & ${2486667}_{-4207}^{+51547}$\\
Mass Ratio & & ${0.080}_{-0.012}^{+0.014}$\\
\enddata
\end{deluxetable}

\section{Atmosphere and Evolution Model Comparison}\label{sec:atmosphereFitting}

In the following sections we show a preliminary comparison of our near-IR photometry from NIRCam with brown dwarf atmospheric models, focusing on the Sonora models \citep{marley2021, karalidi2021}. In our spectral fitting, we only use the model spectral grid with solar carbon-to-oxygen ratio.

The Sonora-Bobcat cloudless atmospheric models assume that the atmospheric composition is in thermo-chemical equilibrium and solve radiative transfer equations for a self-consistent temperature-pressure profile. The model grid covers temperatures from 200 to 2400\,K, gravity from 10 to 3160 $\rm m\, s^{-2}$, and metallicities from [Fe/H]=-0.5 to 0.5.
The model spectra have a spectral resolution ranging from 0.6 to 20\um. 

The Sonora-Cholla cloudless models assume chemical disequilibrium. These models assume that atmospheres have solar metallicity \citep{lodders2010} with cloud-free atmospheric structures. The models are computed using the \texttt{Picaso v3.0} atmospheric model \citep{Mukherjee2022}. By including an eddy diffusion parameter, $K_{\rm zz}$, that ranges from $10^2-10^7\,\rm cm~s^{-2}$ as an input parameter, the Cholla models simulate the dynamical mixing that drives various molecular species like 
CH$_4$, CO, H$_2$O, and NH$_3$
out of their thermochemical equilibrium abundances.
The Cholla models span over a temperature grid of 500 to 1300\,K and a gravity grid from 56 to 3160 $\rm \,cms^{-2}$.

First, we perform a basic $\chi^2$ comparison of new JWST photometry to the Sonora models \citep{marley2021, karalidi2021}. 
We also perform an MCMC fit of the model grids with both our new photometry and previously published medium resolution spectrum obtained with SPHERE-IRDIS LSS \citep{mesa2020} and ground-based photometry from SPHERE-IRDIS \cite{Maire2020}. From this we derive a bolometric luminosity and determine model-dependent mass estimate. 

\subsection{Model Comparison with NIRCam Photometry Only}

We compare the NIRCam photometry with both the Sonora-Bobcat and Sonora-Cholla model grids to highlight the broad features of the $2-5\,\rm \mu m$ flux. We allow the radius scaling to vary when fitting the models to our NIRCam photometry, which is generally consistent with radii much smaller than, e.g., those predicted by the Sonora-Bobcat grid evolutionary tables. 

We find that the Bobcat models do not simultaneously capture both the local flux peak at $3\,\rm \mu m$ and the steep drop in flux from $4-5\,\rm \mu m$. In general the Bobcat grid favors higher gravity and lower or zero metallicity. However, none of the fits capture all of the photometry perfectly. A model spectrum at effective temperature of $T_{eff}=1400K$ best matches the data, but does not fully capture the peak flux at $\sim4\,\rm \mu m$ with the drop off at longer wavelength. Gravity and metallicity do not have a strong effect on the fit.

\begin{figure*}
\begin{center}
\includegraphics[width=0.478\textwidth]{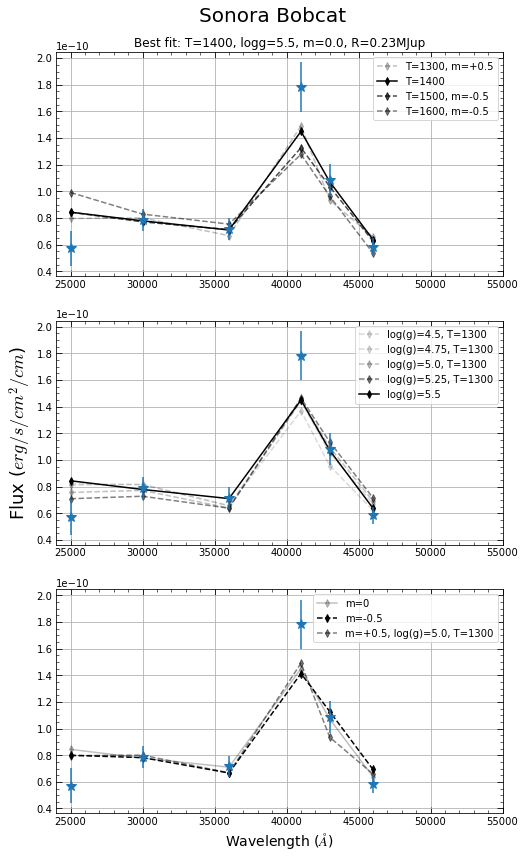}
\includegraphics[width=0.45\textwidth]{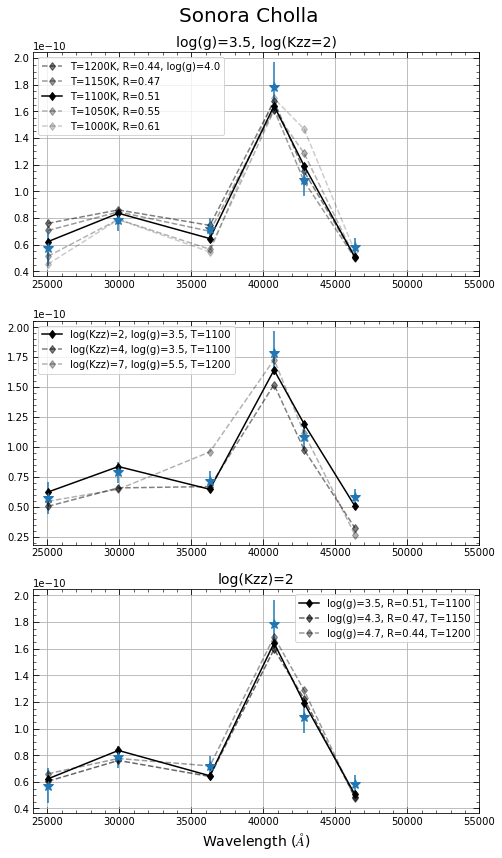} 
\caption{\textbf{Left:} Measured photometry from NIRCam plotted alongside a few closely matching scenarios from the Sonora-Bobcat models, letting the radius vary.
\textbf{Right:} Measured photometry from NIRCam plotted alongside a few closely matching scenarios from the Sonora-Cholla models, which include chemical disequilibrium parameterized by the eddy diffusion parameter, $log(K_{zz})$. \label{fig:sonoraphotfit}}
\end{center}
\end{figure*}

The Sonora Cholla models, which include effects of disequilibrium chemistry, provide better fits to the photometry, but suggest a low value for gravity as well as small radius to scale the flux. The best fit model  has $T_{eff}=1200K$, $g=31$, and $log(K_{zz})=2$. Figure \ref{fig:sonoraphotfit} (right) shows the best fitting Cholla model, with a few model grid points that vary in temperature, eddy diffusion parameter, and gravity. We note that while lower gravity is favored, the gravity does not strongly affect the shape of the photometry-only profile, as can be seen in the third panel. Near-IR spectroscopy with JWST will be able to further distinguished features of gravity. 

Overall, the best fit Sonora-Cholla model provides a better fit than the best fit Sonora-Bobcat model. As can be seen  in Figure \ref{fig:sonoraphotfit}, the Cholla models are better able to simultaneously capture the lower flux at shorter wavelengths, the peak at $\sim$4 \mum, and the subsequent drop in flux, favoring a lower temperature that is more consistent with previous estimates. This suggests that modeling disequilibrium chemistry may be necessary for understanding the atmosphere and evolution of \hdb.

\subsection{Model Fit to 1-5 $\mu m$}

We fit the Sonora-Cholla cloudless models to the composite dataset that consists of JWST photometry, the IRDIS spectra \citep{mesa2020}, and the SPHERE K12 band photometry \citep{Maire2020} to constrain the temperature and gravity of HD19487B.
For the IRDIS spectra, we exclude spectral points with negative flux values.
We include an additional uncertainty of 7\%, similar to the J- and H-band absolute flux uncertainties of \hd\ (Table 7 of \citealt{mesa2020}), to account for the possible offset in the absolute flux levels.

We linearly interpolate the Cholla model spectral grid to a finer grid with smaller step sizes in the temperature, gravity, and eddy diffusion parameter.
In addition to the three parameters, the other free parameter in the spectral fitting is the scaling factor, which is the square of the ratio of brown dwarf radius to the distance (32.03 pc). 
We use \texttt{pyphot \footnote{\url{https://github.com/mfouesneau/pyphot}}} to calculate the broadband photometries of model spectra.

Our model fitting algorithm minimizes a cost function that sums the squared difference between the model spectra and data, weighted by the observational uncertainties, $\sigma$, and the intensities, $w$, as shown in the following:
\begin{multline}
{\mathrm cost~ function} = \Sigma_{i=1}^{N}\{w_i \frac{F_{\lambda,i}(model) - F_{\lambda,i}(data)}{\sigma_i}\}^2 \\
w_i = \frac{F_{\lambda,i} \Delta\lambda_i}{max(F_{\lambda,i} \Delta \lambda_i)}
\end{multline}
where $\Delta \lambda_i$ is the wavelength coverage of a datapoint and the intensity weighting is normalized by the highest intensity among the datapoints.
We include the intensity weighting in the cost function so that a photometric point with high intensity carries more weight in the fitting process than a spectral point with low intensity even though both could have a similar signal-to-noise ratio, since the photometric points contain a larger fraction of total flux measured.
We then use \texttt{emcee} \citep{dfm17} to sample the posterior distribution of the fitted parameters with the Markov Chain Monte Carlo (MCMC) method.
We adopt a uniform prior of temperature, logarithmic gravity, and eddy diffusion parameter.
We run the MCMC chain with 200 walkers for 20,000 steps.
Based on the posterior distribution of the MCMC chains, we derive the best-fit temperature of $1080\pm 22 {\rm \,K}, $ gravity of $\log(g) = 4.60^{+0.2}_{-0.1},$ eddy diffusion parameter of $ \log K_{\rm zz} = 3.1^{+0.6}_{-0.4}$, and radius R of $0.62 \pm 0.03 R_{\rm J}$. The value for radius is lower than the expected radius at 9~Gyr, $\sim$0.8~$\mathrm{R_{Jup}}$, according to the evolution model grid at similar parameters. Prior modeling work has noted a common discrepancy between the expected radius from evolutionary models and the radius required to match the flux of a given temperature that best fits atmospheric models \citep[e.g.,][]{Barman2011,marley2012, Lavie2017}.  
\cite{Maire2020} explored radius agreement with evolutionary tracks with different model atmospheres that included various levels of clouds in the atmosphere. Future observations, especially spectroscopy will further help constrain atmospheric models.

\begin{figure*}
    \centering
    \includegraphics[width=1.1 \textwidth]{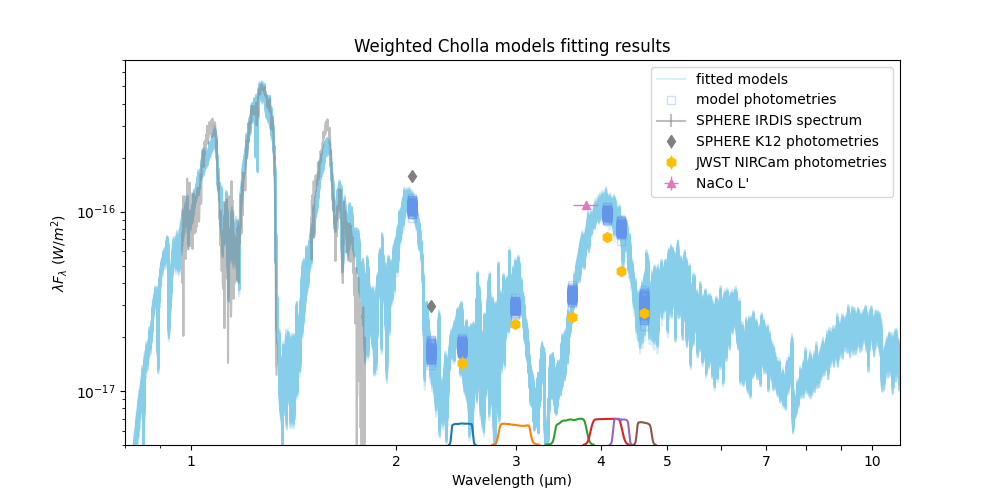}
    \caption{The comparison of the fitted Cholla model spectra (light blue lines) to the SPHERE/IRDIS-LSS 1--1.8\um spectra (gray lines), SPHERE/IRDIS-K12 photometry (gray diamonds), NaCo L' band (pink triangle), and JWST NIRCam photometry (golden hexagon) show that it is challenging for the models to simultaneously explain the near-IR and mid-IR spectra and photometry. The semi-transparent light blues lines are the 1000 Cholla cloudless model spectra sampled from the posterior distribution of MCMC fitting results. The blue squares are the effective photometry from the model spectra samples in the corresponding NIRCam filters. The transmission curves of JWST NIRCam broadband photometry are plotted in colored lines at the bottom of the plot. The y-axis is in unit of intensity ($\rm Wm^{-2}$). \label{fig:lbol}}
\end{figure*}

We draw 100 parameter sets from the posterior distribution and plot the corresponding model spectra in Figure \ref{fig:lbol}. 
Figure \ref{fig:lbol} suggests that the model spectra provide a qualitatively good match to the data but there are some significant residuals, especially in the $K$-band photometries. We remain cautious about the inferred parameters and the seemingly small uncertainties given the imperfect fit between the data and model spectra.
However, while absolute flux calibration may account for some discrepancy between data sources, the discrepancy alone is not enough to account for inconsistencies between the best fit atmospheric model and evolutionary grid predictions.
Clouds, which likely drive the rotational modulation of many T dwarfs \citep[e.g][]{Manjavacas2019} and are not included in the Cholla models, could play a key role in shaping the \hdb\ emission spectra.
Future simultaneous observation in the near-IR and mid-IR region will be useful for testing the role of clouds in the atmosphere of \hdb\ with well-constrained age and host star metallicity.

\subsection{Bolometric Luminosity}

\begin{figure}
    \centering
    \includegraphics[width=0.5\textwidth]{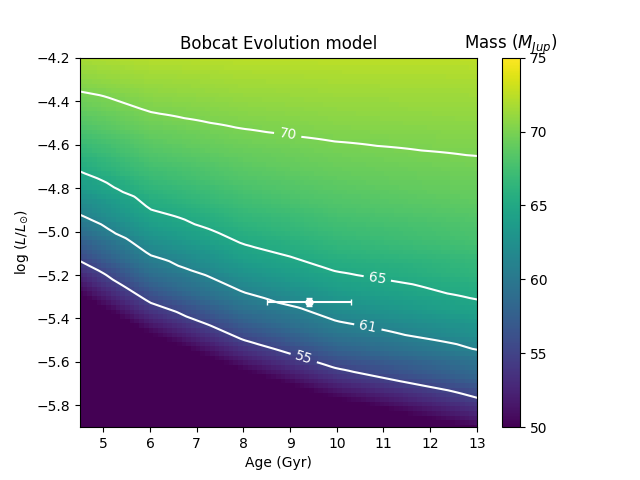}
    \caption{Based on the bolometric luminosity and age, the Bobcat evolution models suggest that \hdb\ has a mass of $62\pm1$ \mj. The colors indicate the masses predicted by Bobcat evolution models. White contour lines show the bolometric luminosity and age with a fixed mass. The uncertainty of \hdb's bolometric luminosity is around 4\%.}
    \label{fig:lbol_evol}
\end{figure}

The 2-5µm JWST NIRCam broadband photometry, in combination with the ground-based near-infrared spectra and photometry, are crucial for pinning down the bolometric luminosity of a $\sim$1000 K object.
We integrate the flux density over  observed data including the previously published IRDIS-LSS spectra (0.97-1.335{\,\mum} \& 1.50-1.80 {\,\mum}) , ground-based K-band photometry (2.059-2.161{\,\mum} \& 2.1965-2.3055{\,\mum}), and the six JWST NIRCam broad band photometry.
The flux integral in the observed wavelength regions is  $3.28 \pm 0.2 \times 10^{-6} L_{\odot}$.

We utilize the fitted model spectra in Section \ref{sec:atmosphereFitting} to extrapolate flux density beyond the observed wavelength region and estimate the bolometric luminosity.
We find that the observational data accounts for around 72\% of the bolometric luminosity.
The estimated total bolometric luminosity is $(4.75 \pm 0.2)\times 10^{-6}L_{\odot}$, or $\log(L/L_{\odot}) = -5.32 \pm 0.02$.
Based on the combination of JWST NIRCam high-precision photometry and ground-based data, our results suggest that we can accurately derive the bolometric luminosity at 4\% precision level.

Based on the independently estimated age and bolometric luminosity, we then use the Bobcat evolution model to estimate mass of \hdb. 
After linearly interpolating the mass as a function of age and bolometric luminosity, we derive that the mass is $62 \pm 1 M_{J}$. Figure \ref{fig:lbol_evol} shows the Sonora Bobcat evolution model and where our bolometric luminosity estimate lies.
By comparing the mass derived from the age and $L_{bol}$ to the dynamical mass, ($81^{+14}_{-12}M_J$), we conclude that the two masses are consistent with each other within about two-sigma.

The NIRSpec IFU observations planned for later this year (PID \#1414) will provide a much more complete characterization of the atmosphere of  \hdb, helping to pin down parameters such as metallicity and \teff.

\section{Conclusions}

We have demonstrated the performance of JWST NIRCam LWB coronagraph on the known binary system \hdb. Despite missing reference star observations, we are able to recover the companion with high significance in all 6 medium NIRCam bands used for the observations. 

The main results of this study are as follows:
\begin{itemize}
    \item The MASKLWB coronagraph works well for separations below $1$~arcsec (for medium filters excluding F250M) at contrasts $10^{-5}$ and better, even without a reference star when angular diversity is utilized. This is expected to improve in the near future when coronagraphic mask locations are refined through instrument calibration observations.
    \item Given the superb stability of JWST, and regular OPD measurements available, we are able to incorporate synthetic reference images to further subtract speckles, following ADI subtraction, and improve SNR on the detections. Future observations with reference observations can be compared with the results presented in this study.
    \item We estimate the age of the HD19467 system by combining spectroscopy and Gaia astrometry with asteroseismic constraints from TESS, finding an age of $9.4\pm1.0$~Gyr, supporting older estimates of the age.  We provide updated parameters for the host star HD19467. 
    \item We estimate a dynamical mass of \hdb\ of $81^{+14}_{-12}M_J$, contributing new relative astrometry from NIRCam and radial velocities from HIRES, and detect tentative evidence of curvature in the orbit fit to the radial velocities. 
    \item A comparison of atmospheric and evolutionary models to our new $2-5$~\mum\ photometry favors models that include disequilibrium chemistry. 
    \item A global fit to the photometry and spectroscopy from this study and ground-based observations show some tension between the instrument-to-instrument relative fluxes and the models. 
    \item The model-derived mass of $62\pm1~M_J$ is lower than the dynamical mass estimate, but within 2-$\sigma$. 
\end{itemize}

The NIRCam observations provide the highest fidelity $3-5$~\mum\ photometry to date of \hdb, and give an early test of atmospheric and evolutionary models that JWST will continue to test throughout the mission. 
Future observations with NIRSpec (PID \#1414) will further elucidate discrepancies in model spectra and help characterize the chemistry of \hdb. Improvements in the near future through instrument calibration observations will further refine the performance of the coronagraphic mask placement and the performance of the MASKLWB mode.

\acknowledgements
The authors thank the anonymous reviewer for helpful comments. The authors acknowledge useful discussions with G. Mirek Brandt and Eric Nielsen. We also thank Dino Mesa for providing the data for the near-IR spectrum of \hdb.

Some of the research described in this publication was carried out at the Jet Propulsion Laboratory, California Institute of Technology, under a contract with the National Aeronautics and Space Administration.
D.J.\ is supported by NRC Canada and by an NSERC Discovery Grant. L.E.U.C.’s research was supported by an appointment to the NASA Postdoctoral Program at the NASA Ames Research Center, administered by Oak Ridge Associated Universities under contract with NASA.
D.R.H.\ and D.H.\ acknowledge support from the Research Corporation for Science Advancement (Scialog award \#26996) and the National Science Foundation (AST-2009828). D.H.\ also acknowledges support from the Alfred P. Sloan Foundation and the National Aeronautics and Space Administration (80NSSC21K0652,80NSSC22K0303).

Some of the data presented in this paper were obtained from the Mikulski Archive for Space Telescopes (MAST) at the Space Telescope Science Institute. The specific observations analyzed can be accessed via \dataset[10.17909/v8m9-xk98]{https://doi.org/10.17909/v8m9-xk98} and \dataset[10.17909/t9-st5g-3177]{https://doi.org/10.17909/t9-st5g-3177}.

This research has made use of the SIMBAD database,
operated at CDS, Strasbourg, France.

Some of the data presented herein were obtained at the W. M. Keck Observatory, which is operated as a scientific partnership among the California Institute of Technology, the University of California and the National Aeronautics and Space Administration. The Observatory was made possible by the generous financial support of the W. M. Keck Foundation.  The authors wish to recognize and acknowledge the very significant cultural role and reverence that the summit of Maunakea has always had within the indigenous Hawaiian community.  We are most fortunate to have the opportunity to conduct observations from this mountain.

\software{SciPy \citep{2020SciPy-NMeth}, NumPy \citep{harris2020array}, Matplotlib \citep{matplotlib}, WebbPSF \citep{Perrin2014}, PyKLIP \citep{Wang2015}, synphot \citep{synphot}, pysynphot \citep{2013ascl.soft03023S}, pyphot \citep{pyphot}
}

\appendix

\section{KLIP Forward Model of the Data} \label{sec:fmstamps}
\begin{figure}[b!]
    \centering
    \includegraphics[width=.32\textwidth]{fm_residuals_stamp_F250M.png}
    \includegraphics[width=.32\textwidth]{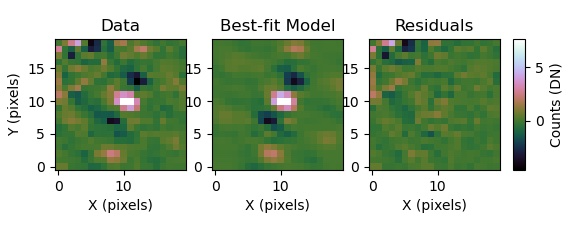}
    \includegraphics[width=.32\textwidth]{fm_residuals_stamp_F360M.png}
    \includegraphics[width=.32\textwidth]{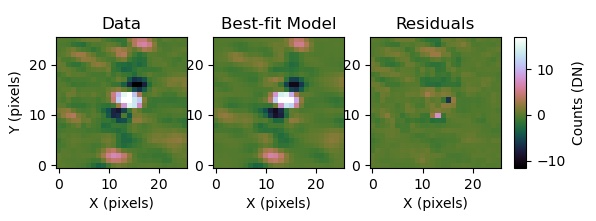}
    \includegraphics[width=.32\textwidth]{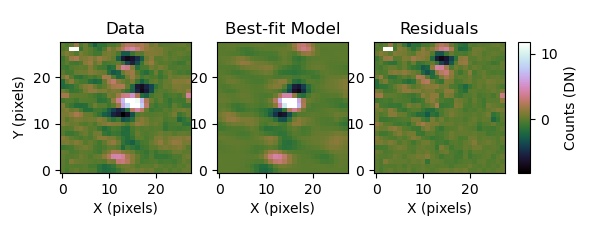}
    \includegraphics[width=.32\textwidth]{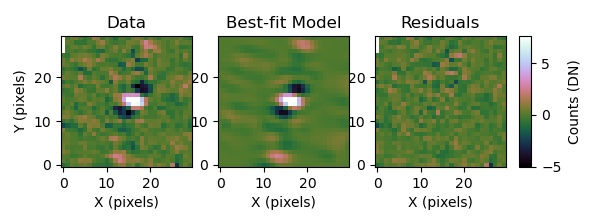}
    \caption{The KLIP forward model that accounts for over subtraction effects compared to the PSF-subtracted data in each of size filters. }
    \label{fig:fmcomparison_all}
\end{figure}

We compute a forward model of the PSF-subtracted image according to \citet{Pueyo2016} to account for over- and self-subtraction effects. ADI imaging is particularly susceptible to self-subtraction effects, especially when there is little angular diversity, such as our data, which contains only two roll angles separated at $\sim 8$ degrees. Modeling these effects is essential for appropriately estimating the properties of the signal (flux, position). Figure \ref{fig:fmcomparison_all} displays the comparison between the forward model and the PSF-subtracted image corresponding to the images displayed in Figure \ref{fig:NIRcamPCA}. 

\section{Keplerian Orbit Fit to HD 19467 Radial Velocities}
\label{sec:RVapp}
We fit a Keplerian Orbit model to radial velocities of \hd ~from HIRES and HARPS instruments, the latter published in \cite{trifonov2020}, catalog J\/A+A\/636\/A74\/rvbank table entries DRVmlcnzp and e\_DRVmlcnzp.
HIRES radial velocities, along with the new measurements are presented in Table \ref{tab:NEW_RV}. Using \texttt{RadVel} \citep{fulton2018}, we determine that a model with a curvature term is slightly favored over a model without curvature, however the trend-only and trend plus curvature models are nearly indistinguishable (Table \ref{tab:modelcomp}), based on $\Delta BIC$ and $\Delta AIC$ metrics. This is the first tentative evidence of curvature measured for \hd. Future follow up is needed to further support this finding.

\begin{deluxetable}{llll}
\centering
\tablecaption{HIRES Radial Velocity Observations\label{tab:NEW_RV}}
\tablehead{
\colhead{Instrument} & \colhead{BJD$_{TDB}$}& \colhead{RV} & \colhead{RV Error}}
\startdata
HIRES & 2450366.019	& 20.64 &	1.47 \\
HIRES & 2450418.943	& 29.50  &	2.17 \\
HIRES & 2450461.84	& 34.01  &	1.30 \\
HIRES & 2450715.103	& 26.13  &	4.20 \\
HIRES & 2450716.111	& 25.36  &	4.31 \\
HIRES & 2450786.847	& 23.29  &	2.38 \\
HIRES & 2450786.86	& 27.42 &	1.51 \\
HIRES & 2450806.904	& 22.50  &	1.50 \\
HIRES & 2450837.744	& 14.19 &	1.41 \\
HIRES & 2450839.743	& 19.41 &	1.49 \\
HIRES & 2451012.119	& 27.88 &	1.40 \\
HIRES & 2451013.12	& 19.74 &	1.31 \\
HIRES & 2451070.116	& 20.60 &	4.26 \\
HIRES & 2451072.984	& 29.72  &	4.36 \\
HIRES & 2451171.777	& 26.06  &	1.48 \\
HIRES & 2451410.128	& 21.00  &	1.51 \\
HIRES & 2451543.847	& 18.48  &	1.60 \\
HIRES & 2451551.792	& 19.77 &	1.47 \\
HIRES & 2451552.844	& 14.65 &	1.73 \\
HIRES & 2451582.73	& 24.84 &	1.70 \\
HIRES & 2451882.806	& 29.84 &	1.62 \\
HIRES & 2451900.783	& 23.36  &	1.48 \\
HIRES & 2452134.08	& 20.03 &	1.58\\
HIRES & 2452242.908	& 19.50 &	1.42\\
HIRES & 2452516.022	& 18.34 &	1.62 \\
HIRES & 2452575.902	& 10.04 &	1.72 \\
HIRES & 2452835.128	& 19.96 &	1.84 \\
HIRES & 2452926.089	& 19.92 &	4.43 \\
HIRES & 2453240.043	& 11.43 &	1.16 \\
HIRES & 2453427.785	& 8.38 &	1.20 \\
HIRES & 2453984.039	& 5.01 &	1.08 \\
HIRES & 2455807.035	& -3.65 &	1.23 \\
HIRES & 2455808.105	& -0.48 &	 1.42 \\
HIRES & 2455809.088	& 1.84 &	1.22 \\
HIRES & 2455903.779	& 8.54 &	1.33 \\
HIRES & 2456152.11	& -2.44 &	1.18 \\
HIRES & 2456210.015	& 1.14 &	1.49 \\
HIRES & 2456519.085	& -0.97 &	1.28 \\
HIRES & 2456530.025	& -7.90  &	1.20 \\
HIRES & 2456548.035	& -0.47 &	1.33 \\
HIRES & 2456586.036	& -2.87 &	1.41 \\
HIRES & 2456587.965	& -3.16 &	1.42 \\
HIRES & 2456588.997	& 4.29  &	1.41 \\
HIRES & 2456613.908	& -2.82 &	1.38 \\
HIRES & 2456637.791	& -0.19 &	1.45 \\
\hline
\multicolumn{4}{c}{\textit{New data}} \\
HIRES & 2457245.143	& -0.38 & 1.18 \\
HIRES & 2458367.035	& -15.44 & 1.76 \\
HIRES & 2459632.706	& -10.11 & 1.51 \\
HIRES & 2459649.719	& -13.26 & 1.48 \\
HIRES & 2459780.128	& -14.46 & 1.29 \\
HIRES & 2459786.103	& -4.38 & 1.16 \\
HIRES & 2459787.129	& -15.52 & 1.20 \\
\enddata
\end{deluxetable}

\begin{deluxetable}{llrrrrrrr}
\tablecaption{Model Comparison \label{tab:modelcomp}}
\tablehead{\colhead{AICc Qualitative Comparison} & \colhead{Free Parameters} & \colhead{$N_{\rm free}$} & \colhead{$N_{\rm data}$} & \colhead{RMS} & \colhead{$\ln{\mathcal{L}}$} & \colhead{BIC} & \colhead{AICc} & \colhead{$\Delta$AICc}}
\startdata
  AICc Favored Model & $\dot{\gamma}$, $\ddot{\gamma}$, {$\sigma$}, {$\gamma$} & 8 & 128 & 3.40 & -330.13 & 695.02 & 673.41 & 0.00 \\
  \hline \\
  Nearly Indistinguishable & $\dot{\gamma}$, {$\sigma$}, {$\gamma$} & 7 & 128 & 3.40 & -332.43 & 692.91 & 673.88 & 0.47 \\
  \hline \\
  Ruled Out & {$\sigma$}, {$\gamma$} & 6 & 128 & 9.50 & -426.33 & 872.85 & 856.43 & 183.02 \\
\enddata
\end{deluxetable}

\section{The MCMC posterior distribution for spectral fitting}\label{sec:mcmc_cholla}
\begin{figure*}
    \centering
    \includegraphics[width=.6\textwidth]{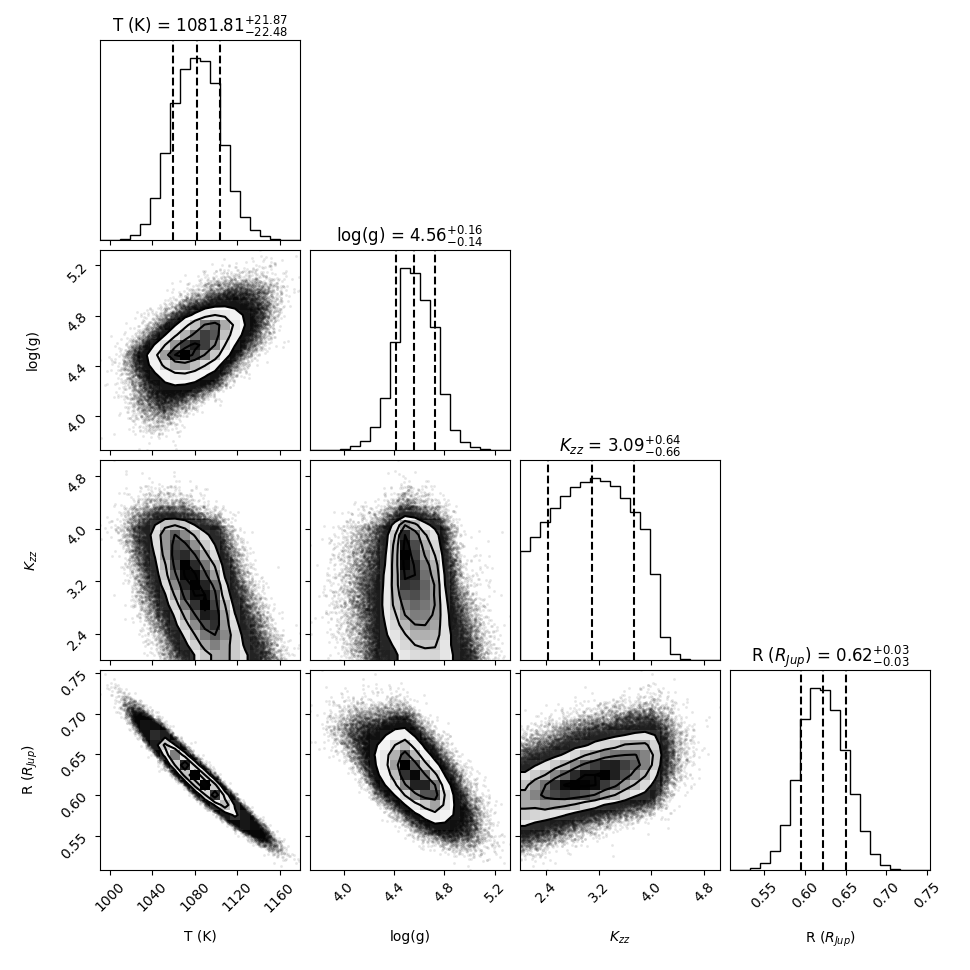}
    \caption{The posterior distribution of the MCMC fitting of Cholla models to the SPHERE-IRDIS spectra, SPHERE K12 photometry, and the JWST NIRCam photometries.}
    \label{fig:corner_cholla}
\end{figure*}

Figure \ref{fig:corner_cholla} shows the MCMC posterior distribution of the spectral fitting. The spatial structure seen in the posterior distribution reflects the step size of temperature (10\,K), gravity (0.025 dex), and $\log(K_{\rm zz})$ (0.167) in the interpolated model grid. The sharp boundaries of radius at 0.5 $\rm R_{Jup}$ is reflects the lowest limit of radius range (0.5-1.5 $\rm {R_{Jup}}$) in the model fitting.

\clearpage
\bibliography{ms.bib}

\end{document}

%% file: commands-star.tex
\newcommand{\logg}{\mbox{$\log g$}}
\newcommand{\muHz}{\mbox{$\mu$Hz}}
\newcommand{\mh}{\mbox{$\rm{[M/H]}$}}

\newcommand{\numax}{\ensuremath{\nu_{\textrm{max}}}}
\newcommand{\dnu}{\ensuremath{\Delta\nu}}

\newcommand{\rhostar}{\mbox{$\rho_{\star}$}}
\newcommand{\rstar}{\mbox{$R_{\star}$}}

\newcommand{\mstar}{\mbox{$M_{\star}$}}

\providecommand{\mj}{\ensuremath{\,M_{\rm J}}}

\newcommand{\teffstar}{\mbox{$5747 \pm 40$}}
\newcommand{\fehstar}{\mbox{$-0.09 \pm 0.04$}}
\newcommand{\radstar}{\mbox{$1.20\pm0.03$}}
\newcommand{\lumstar}{\mbox{$1.42\pm0.06$}}
\newcommand{\massstar}{\mbox{$0.96\pm0.02$}}
\newcommand{\agestar}{\mbox{$9.4\pm1.0$}}
\newcommand{\loggstar}{\mbox{$4.28\pm0.04$}}
\newcommand{\denstar}{\mbox{$0.56\pm0.03$}}